\documentclass[11pt,fleqn]{article}  \makeatletter
\newlength{\skbaseline} 
\usepackage{makeidx}    \makeindex
\usepackage{fancyhdr}   \usepackage{fancybox}  \usepackage{wrapfig}
\usepackage[dvips]{graphics}
\usepackage{amsmath,amsbsy,amssymb}
\usepackage[mathscr]{eucal}
\usepackage{amsfonts}
\begin{document}

\newcommand{\bra}[1]{\langle \, #1 \, |} 
\newcommand{\ddt}{\frac{d}{dt}\:} 
\newcommand{\elm}[3]{\bra{#1}\,#2\,\ket{#3}} 
\newcommand{\idd}{\mathbf{1}} 
\newcommand{\inro}{\widetilde{\varrho}} 
\newcommand{\ket}[1]{| \, #1 \, \rangle} 
\newcommand{\pol}{{\textstyle \frac{1}{2}}} 
\newcommand{\qav}[1]{\langle\,#1\,\rangle} 
\newcommand{\sib}{\bar{\sigma}_{B}} 
\newcommand{\TTr}{\mathrm{Tr}} 

\headheight 16pt 
\pagestyle{fancy}     \headwidth 140mm
\renewcommand{\headrulewidth}{0.5pt}
\renewcommand{\footrulewidth}{0.5pt}

\lhead[\footnotesize \sf ME -- Tutorial]
      {\footnotesize \sf ME -- Tutorial}
\rhead[]{\bf \thepage}
\chead[\footnotesize \sf \leftmark]{\footnotesize \sf \leftmark}
\cfoot[]{}
\lfoot[\footnotesize \sf S.K, J.C-K]{\footnotesize \sf S.K, J.C-K}
\rfoot[\bf \thepage]{\bf \thepage}

\clearpage{\pagestyle{empty}\cleardoublepage}

\title{\bf Master equation -- tutorial approach}
\author{Stanis{\l}aw Kryszewski\thanks{Email: fizsk@univ.gda.pl}
~and Justyna Czechowska-Kryszk\\
Institute of Theoretical Physics and Astrophysics, \\
University of Gda{\'{n}}sk, ul. Wita Stwosza 57,
80-952 Gda{\'{n}}sk, POLAND}
\maketitle

\begin{abstract}
We do not present any original or new material. This is a
tutorial addressed to students who need to study the
microscopic derivation of the quantum-mechanical
master equation encountered in many practical physical
situations.
\end{abstract}

\renewcommand{\thepage}{\arabic{page}}   \setcounter{page}{0}
\footnotesize \parskip=-1mm
\setlength{\baselineskip}{0.75\skbaseline}
\tableofcontents \thispagestyle{empty} 
\normalsize

\setlength{\baselineskip}{\skbaseline}

\newpage

\setlength{\baselineskip}{\skbaseline}
\section{Introduction}

\subsection{Motivation}

The term "master equation" (ME) used in this work means an
equation of motion for the reduced density operator $\rho_{A}(t)$
of subsystem $\cal{A}$ which interacts with (usually much larger)
another subsystem $\cal{B}$. Experiences gathered during quite
a few years of lecturing indicate that students find it difficult
to understand the derivation and consequences of ME.
The literature sources (at least those known to us) are
usually quite brief and not easy to follow.

The aim of this paper is to give a full presentation of the
so-called microscopic derivation of ME together with detailed
discussion of the underlying assumptions and approximations.
Hence, this work does not bring any original material.
It is just a tutorial which, hopefully, will help the students
to obtain a better grasp of the ideas and concepts leading to
one of important theoretical methods used in a variety
of physical problems.

The literature of the subject is quite rich. Entering the term
"master equation" into {\em Google Scholar} returns about 30 000
links. It seems virtually impossible to give an extensive
bibliography. Therefore, we concentrate only on several works
which were essential in the preparation of this paper.

\subsection{Outline of the problem}

In order to present the ME technique, we first give the basic
ideas which help, so to speak, to set the scene for further
developments. We assume that the reader is familiar with
fundamental concepts of quantum mechanics (given, for example,
in the main chapters of the first volume of the excellent
book by Cohen-Tannoudji {\em et al} \cite{coh0}).

To start with, we recall that the state of the quantum system is,
in the majority of practical applications, given by the
density operator. This concept is introduced and discussed in
virtually all handbooks on quantum mechanics,
hence we mention only the basic facts which will be needed here.
Namely, density operator for any physical system must have three
essential properties
\begin{subequations} \label{ii01}
\begin{alignat}{2}
  \rho &= \rho^{\dagger}, &\quad - \quad & \mathrm{hermiticity};
\label{ii01a} \\
  \TTr\{ \rho \} &= 1, &\quad - \quad & \mathrm{normalization};
\label{ii01b} \\
  \rho &\ge 0, &\quad - \quad & \mathrm{semi-positivity}.
\label{ii01c} \end{alignat}
\end{subequations}
The last inequality means that $\elm{\psi}{\rho}{\psi} \ge 0$
for any state $\ket{\psi}$ from the Hilbert space of the states
of the given physical system. These properties may be phrased
in terms of the eigenvalues $\lambda_{k}$ of the density
operator. Namely, $\lambda_{k} \in \mathbb{R}$,
$\sum_{k} \lambda_{k} = 1$, and $\lambda_{k} \ge 0$.

For a closed system with hamiltonian $H$, the evolution of the
corresponding density operator is given by the von Neumann
equation
\begin{equation}
  i\hbar \: \ddt \: \rho(t) = \bigl[ \: H, \: \rho(t) \: \bigr],
\label{ii03} \end{equation}
which has a well-known solution
\begin{equation}
   \rho(t)
   = U(t,t_{0}) \: \rho(t_{0}) \: U^{\dagger}(t,t_{0}),
     \quad\mathrm{with}\quad
     U(t,t_{0}) = \exp \left( -\frac{i}{\hbar} H(t-t_{0}) \right),
\label{ii04} \end{equation}
where $\rho(t_{0})$ is a suitably chosen initial condition.
Such an evolution is unitary \cite{coh0} and obviously
preserves all the necessary properties \eqref{ii01} of
the density operator. In such a case everything is conceptually
clear although necessary calculation may be quite involved
or even requiring some approximate computational methods.

The problem arises when we deal with a bipartite system
$\cal{A} + \cal{B}$, consisting of two interacting subsystems
$\cal{A}$ and $\cal{B}$. Let us briefly outline the physical
situation. We assume that the whole system $\cal{A} + \cal{B}$
is closed and the total hamiltonian is written as
\begin{equation}
      H_{AB} = H_{0} + V_{AB},
      \qquad \mathrm{where} \qquad
      H_{0} ~=~ H_{A}\otimes\idd_{B} ~+~\idd_{A}\otimes H_{B},
\label{ii05} \end{equation}
where $H_{A}$ and $H_{B}$ are free, independent hamiltonians
of each of the subsystems ${\cal A}$ and ${\cal B}$.
$V_{AB}$~is the hamiltonian describing the interaction
between two parts. Let $\rho_{AB}(t)$ denote  the density
operator of the total system. Then, $\rho_{AB}(t)$
evolves according to the von Neumann equation
(identical as \eqref{ii03})
\begin{equation}
  i\hbar \: \ddt \: \rho_{AB}(t)
    = \bigl[ \: H_{AB}, \: \rho_{AB}(t) \: \bigr],
\label{ii06} \end{equation}
Now, one may ask, what is the problem? The point is that
we are, in fact, interested only in the subsystem $\cal{A}$.
Subsystem $\cal{B}$, for this reason or other, is considered
irrelevant, although the $\cal{A}$-$\cal{B}$ interaction
certainly affects the evolution of $\cal{A}$. Moreover,
in many practical cases, subsystem $\cal{B}$ is much larger
(with many more degrees of freedom) and virtually inaccessible
to direct measurements. Frequently, $\cal{B}$ is a reservoir
\cite{al,bp} and plays the role of environment upon which we have
neither control nor influence. This may be very important
in the context of quantum information theory \cite{horn,pres}
when the relevant subsystem is disturbed by the surroundings.
Moreover, problems of decoherence and irreversibility are
intrinsically connected with the effects occurring in a subsystem
influenced by an external reservoir (see \cite{bp,horn}).
We shall not discuss these problems but focus on master
equation technique.

In the view of these brief remarks the question is, how to extract
useful information on $\cal{A}$ from the general von Neumann
equation \eqref{ii06}. We stress that we are interested only in
system $\cal{A}$, so we need to find the reduced density operator
\begin{equation}
   \rho_{A}(t) = \TTr_{B} \{ \rho_{AB}(t) \}.
\label{ii07} \end{equation}
The aim is, therefore, twofold:
\begin{itemize}
\item[{\em (i)}] extract the evolution of $\rho_{A}(t)$ from
     Eq. \eqref{ii06} for the entire system $\cal{A}+\cal{B}$.
\item[{\em (ii)}] do it in a way which guarantees that properties
     \eqref{ii01} of $\rho_{A}(t)$ (as of any density operator)
     are preserved for any moment of time.
\end{itemize}

The solution to the stated problems is found in the so-called
master equation technique. There are two, conceptually
different but complementary, approaches to ME.

The first one uses mathematically rigorous methods. Such an
approach is presented, for example, in Refs. \cite{al,bp}
(see also the references given in these books).
Rigorous mathematics is obviously very important from the
fundamental point of view. Mathematical theorems prove that
ME for the reduced density operator of subsystem $\cal{A}$
follows from general von Neumann equation \eqref{ii06} and
indeed preserves properties \eqref{ii01}.
In Refs. \cite{al,bp,horn} it is shown that it is so, when ME
attains the so-called standard
(Lindblad-Gorini-Kossakowski-Sudarshan) form
(Ref. \cite{al}, p.8, Eq. (38); \cite{bp}, Eq. (3.63)
and \cite{horn}, Eq. (78))
\begin{equation}
   \ddt \rho_{A}(t)
   = \frac{1}{i\hbar} \bigl[ H, \; \rho_{A}(t) \bigr]
   ~+~ \sum_{ij} a_{ij}
       \biggl( \, F_{i} \, \rho_{A}(t) \, F_{j}^{\dagger}
        ~-~ \pol \bigl[ F_{j}^{\dagger} F_{i}, \; \rho_{A}(t)
                  \bigr]_{+}
       \, \biggr).
\label{ii10} \end{equation}
The first term leads the unitary (hamiltonian) evolution,
while the second one is sometimes called a dissipator \cite{bp}.
Operators $F_{i}$ constitute a basis in the space
of operators for $\cal{A}$ subsystem (see \cite{al,bp}).
Finally, $a_{ij}$ is a positive definite hermitian matrix.
Here, we do not go into the details of the derivation
of the standard form \eqref{ii10} of the master equation,
we refer the reader to the mentioned references
(see also \cite{horn}). Quite interesting derivation of standard
ME is given by Preskill \cite{pres}. His presentation is
somewhat heuristic but certainly worth reading, especially
if one is interested in connection with quantum information
theory, quantum channels etc. It is not, however, our aim
to pursue formal mathematical issues. Our intentions are
quite practical, so the reader may ask, why do we speak about
the standard form of ME. The reason is as follows.

The second approach to the stated problems is via "microscopic
derivations". This occurs when we need to consider a specific
physical situation when interacting systems $\cal{A}$ and $\cal{B}$
are known and well-defined. Then, we want to construct the
corresponding ME -- equation of motion for the reduced density
operator $\rho_{A}(t)$. This is less formal and may be
mathematically uncertain. If the microscopic derivation yields
an equation in the standard form \eqref{ii10}, we can say that
the aim is achieved, because standard form ensures the preservation
of the necessary properties of reduced density operator $\rho_{A}(t)$.
Hence, both approaches are complementary. Formal
but rigorous mathematical methods lead to standard form of ME
which must be matched by equations obtained through
microscopic derivation.

All of the already mentioned references give (usually brief)
account on the microscopic derivation of ME. These presentations
seem to be difficult for students who are not acquainted with
the subject and who seek the necessary introduction.
Perhaps the most extensive microscopic derivation is given
by Cohen-Tannoudji, Dupont-Roc and Grynberg \cite{coh}.
This latter presentation is somewhat heuristic and, as it seems
to us, leaves some nuances unexplained. There is, however,
one more drawback. Namely, Cohen-Tannoudji {\em et al}
do not compare their ME with the  standard form. Therefore,
essential question of positivity preservation remains
untouched.

The derivation given here uses the concepts which can be found
in the all references. Nevertheless, we are most indebted
to Cohen-Tannoudji {\em et al}. Their approach and especially
their discussion of the time scales and employed approximations
strongly influenced our tutorial. We apologize if, at some places,
we do not give proper references. Too many ones can distract
the reader, which does not lessen our debt to all authors
of cited literature.

The scheme of this paper is summarized in the {\em Contents},
hence we feel no need to repeat it. We try to be as clear and
as precise as possible. We focus our attention on the microscopic
derivation of ME, leading to the standard form. Some
issues are postponed to {\em Auxiliary} section, so that
the main flow of derivation is not broken by additional
comments, which can safely be given later.
We hope, that the students who need a primer
in the subject, will find our work useful and informative.
The readers are invited to comment, so that the next version
(if such a need arises) will be improved and, more readable.

\section{Evolution of the reduced density operator}

\subsection{Introductory remarks}

We consider a physical system which consists of two parts
$\cal{A}$ and $\cal{B}$. We are interested only in what happens
in part $\cal{A}$ which is usually much smaller than part
$\cal{B}$. The latter one we will call a reservoir (environment).
We assume that the entire system $\cal{A}+\cal{B}$ is closed.
Then, its hamiltonian is specified as in Eq. \eqref{ii05}
Some additional assumptions concerning both subsystems will be
introduced when necessary.

A previously, let $\rho_{AB}(t)$ denote the density operator
of the whole system  $\cal{A}+\cal{B}$. The evolution of this
operator is governed by von Neumann equation \eqref{ii06}.
Our main aim is to find the corresponding equation of motion
for the reduced density operator $\rho_{A}=\TTr\{ \rho_{AB} \}$
for the subsystem $\cal{A}$. Our starting point is provided
by von Neumann equation which, after the transformation
to the interaction picture, reads
\begin{equation}
  \ddt \: \inro_{AB}(t)
    ~=~ \frac{1}{i\hbar}
       ~\bigl[ \: \widetilde{V}_{AB}(t), \: \inro_{AB}(t) \:
        \bigr],
\label{me05} \end{equation}
where, we obviously denoted
\begin{equation}
   \inro_{AB}(t)
   ~=~ e^{ iH_{0}t/\hbar} \; \rho_{AB}(t) \; e^{-iH_{0}t/\hbar},
   \hspace*{7mm}
   \widetilde{V}_{AB}(t)
   ~=~ e^{ iH_{0}t/\hbar} \; V_{AB} \; e^{-iH_{0}t/\hbar},
\label{me06} \end{equation}
with $H_{0}$ given in Eq. \eqref{ii05}. Reduction of the density
operator (as in \eqref{ii07}) is preserved in the interaction
picture (see {\em Auxiliary} sections)
\begin{equation}
   \inro_{A}(t) ~=~ \TTr_{B}\{ \inro_{AB}(t) \}.
\label{me06c} \end{equation}
Formal integration of Eq. \eqref{me05} yields the following
expression
\begin{equation}
  \inro_{AB}(t+\Delta t)
    ~=~ \inro_{AB}(t)
    ~+~ \frac{1}{i\hbar} \int\limits_{t}^{t+\Delta t} dt_{1}
     ~\bigl[ \: \widetilde{V}_{AB}(t_{1}), \:
             \inro_{AB}(t_{1}) \:
      \bigr],
\label{me07} \end{equation}
which gives the density operator at a later moment  $t+\Delta t$,
while the initial one at a moment $t$ is assumed to be known.
Iterating further and denoting
\begin{equation}
   \Delta \inro_{AB}(t)
   ~=~ \inro_{AB}(t+\Delta t) ~-~ \inro_{AB}(t),
\label{me10} \end{equation}
we obtain, (similarly as in Ref.\cite{coh})
\begin{align}
& \hspace*{-8mm}
   \Delta \inro_{AB}(t)
   ~=~ \left( \frac{1}{i\hbar} \right)
       \int\limits_{t}^{t+\Delta t} dt_{1}
      ~\bigl[ \: \widetilde{V}_{AB}(t_{1}), \: \inro_{AB}(t) \:
       \bigr]
\nonumber \\
& \hspace*{8mm}
    ~+~ \left( \frac{1}{i\hbar} \right)^{2}
        \int\limits_{t}^{t+\Delta t} dt_{1}
        \int_{t}^{t_{1}} dt_{2}
       ~\bigl[ \: \widetilde{V}_{AB}(t_{1}), \;
        \bigl[ \: \widetilde{V}_{AB}(t_{2}), ~\inro_{AB}(t) \:
        \bigr]
\nonumber \\[2mm]
& \hspace*{-8mm}
    ~+~ \left( \frac{1}{i\hbar} \right)^{3}
        \int\limits_{t}^{t+\Delta t} dt_{1}
        \int_{t}^{t_{1}} dt_{2} \int_{t}^{t_{2}} dt_{3}
       ~\bigl[ \: \widetilde{V}_{AB}(t_{1}), \;
        \bigl[ \: \widetilde{V}_{AB}(t_{2}),
        \bigl[ \: \widetilde{V}_{AB}(t_{3})~\inro_{AB}(t_{3}) \:
        \bigr].
\label{me12} \end{align}
Higher order iterations will contain fourfold, etc., integrals and
commutators. Let us note that in the last term we have time ordering
$t+\Delta t \ge t_{1} \ge t_{2} \ge t_{3} \ge t$. The above equation
is rigorous, no approximations have been made.

\subsection{Weak-coupling approximation}

Weak-coupling approximation (discussed in all
Refs.\cite{al}--\cite{coh}), consists in retaining the terms
up to the second order in interaction hamiltonian. Higher order
terms are then neglected. Thus, we remain with
\begin{align}
& \hspace*{-4mm}
  \Delta \inro_{AB}(t)
   ~=~ \left( \frac{1}{i\hbar} \right)
       \int\limits_{t}^{t+\Delta t} dt_{1}
      ~\bigl[ \: \widetilde{V}_{AB}(t_{1}), \: \inro_{AB}(t) \:
       \bigr]
\nonumber \\
& \hspace*{14mm}
    ~+ \left( \frac{1}{i\hbar} \right)^{2}
       \int\limits_{t}^{t+\Delta t} dt_{1}
       \int_{t}^{t_{1}} dt_{2}
      ~\bigl[ \: \widetilde{V}_{AB}(t_{1}), \;
       \bigl[ \: \widetilde{V}_{AB}(t_{2}), ~\inro_{AB}(t) \:
       \bigr].
\label{me16} \end{align}
Alternatively, we can say that the obtained equation is valid
in the second-order perturbation theory. Such an approximation
requires a justification. This will be presented in the
{\em Auxiliary} sections, now we focus on further
steps of the derivation.

Reduction of the operator $\inro_{AB}(t)$ in the left hand
side of \eqref{me16} poses no difficulties.
Tracing over the reservoir variables (subsystem ${\cal B}$)
we obtain
\begin{align}
& \hspace*{-4mm}
  \Delta \inro_{A}(t)
   ~=~ \left( \frac{1}{i\hbar} \right)
       \int\limits_{t}^{t+\Delta t} dt_{1}
      ~\TTr_{B} \bigl[ \: \widetilde{V}_{AB}(t_{1}), \:
       \inro_{AB}(t) \:
       \bigr]
\nonumber \\
& \hspace*{14mm}
    ~+~ \left( \frac{1}{i\hbar} \right)^{2}
        \int\limits_{t}^{t+\Delta t} dt_{1}
        \int_{t}^{t_{1}} dt_{2} ~\TTr_{B}
      \bigl[ \: \widetilde{V}_{AB}(t_{1}), \;
      \bigl[ \: \widetilde{V}_{AB}(t_{2}), ~\inro_{AB}(t) \:
      \bigr].
\label{me16x} \end{align}
This expression has certain drawback. The the commutators in the
right hand side contain full density operator  $\inro_{AB}(t)$,
and not the  interesting (relevant) reduced one $\inro_{A}(t)$.
To proceed, we need some more assumptions and approximations.

One more remark seems to be in place. Subsequent iterations
leading to Eq. \eqref{me12} are rigorous.
In equation \eqref{me16x} -- which is approximate --
there occurs the operator $\rho_{AB}(t)$, taken at the initial
moment. The last term in the exact equation \eqref{me12}
contains $\inro_{AB}$ for moments earlier than the current moment
$t+\Delta t$, but later than the initial instant $t$.
This means that we neglect the influence of the "history"
on the present moment. We shall return to the discussion
of this point.

\subsection{Neglecting the initial correlations}

The key role in our consideration is played by the assumption
that there are two distinct time scales \cite{coh}.
The first one is specified
by $\tau_{B}$ -- typical time during which the internal
correlations in the reservoir $\cal{B}$ exist. This will discussed
in more detail later.  Here we will only say that $\tau_{B}$
is such a time, that when it elapses, the state of the reservoir
is practically independent of its initial  state. The second
scale is provide by time  $T_{A}$. It characterizes the
evolution (changes) of the operator $\inro_{A}(t)$ which is (are)
due to the interaction with reservoir, and which may be specified
by the relation
\begin{equation}
   \frac{\Delta \inro_{A}(t)}{\Delta t}
   ~\sim~ \frac{1}{T_{A}} \; \inro_{A}(t).
\label{me17} \end{equation}
Time $T_{A}$ may be called the characteristic relaxation
time of subsystem ${\cal A}$. Let us note that we are
speaking about interaction -- the interaction picture we employ
is thus, particularly useful. We make no statements about
the rate of the free evolution of $\rho_{A}$
(in the Schr{\"{o}}dinger picture), which is governed by
hamiltonian  $H_{A}$. Usually, the characteristic times of free
evolution (the times of the order of
$\tau_{A} \sim \qav{H_{A}}_{A}/\hbar$) are typically much shorter
that $T_{A}$ describing the influence of the interaction between
subsystems \cite{coh}.

Now, we assume that the introduced time scales satisfy the requirement
\begin{equation}
   \tau_{B} \ll \Delta t \ll T_{A}.
\label{me18} \end{equation}
We have a fast scale (small $\tau_{B}$) determining the decay
of correlations within the reservoir and the second -- much
slower -- scale defined by relatively long relaxation time
$T_{A}$, characterizing the interaction between the two parts
of the entire physical system. This may be phrased differently.
We have assumed that the interaction is weak. Let $V$ denote the
average "strength" of this interaction. Uncertainty principle
states that
\begin{equation}
   VT_{A} \sim \hbar
   \quad \Longrightarrow \quad
   T_{A} \sim \frac{\hbar}{V}.
\label{me19} \end{equation}
The condition $\tau_{B} \ll T_{A}$ implies that
\begin{equation}
   \tau_{B} \ll T_{A} \sim \frac{\hbar}{V}
   \quad \Longrightarrow \quad
   \frac{V \tau_{B}}{\hbar} \ll 1.
\label{me20} \end{equation}
In other words, we can say that spectral widths are the
reciprocals of characteristic times, so the condition
$\tau_{B} \ll T_{A}$ means that the spectral width of the reservoir
energies must be much larger than the spectral width of the
interaction between subsystem ${\cal A}$ with reservoir. Further
discussion and justification of our approximations is postponed
to {\em Auxiliary} sections. Here we focus on the derivation of the
master equation.

The adopted assumption $\tau_{B} \ll T_{A}$ allows us to make
the approximation which is sometimes called the Born one
\cite{bp}-\cite{coh}. Initial density operator for the whole
system ${\cal A} + {\cal B}$ can always be written as
\begin{equation}
  \inro_{AB}(t) ~=~ \inro_{A}(t) \otimes \inro_{B}(t)
      ~+~ \inro_{corel}(t),
\label{me21} \end{equation}
where $\inro_{A}(t)$ and $\inro_{B}(t)$ are the reduced
density operators for two subsystems. The state of the whole system
consists of a factorizable part $\inro_{A}(t) \otimes \inro_{B}(t)$
and the entangled part $\inro_{corel}(t)$, describing the
interaction-induced correlations between the subsystems.
Equation \eqref{me16x} gives us the change
$\Delta \inro_{A}(t) = \inro_{A}(t+\Delta t) - \inro_{A}(t)$,
hence informs us about changes occurring in the time interval
$\Delta t$. Assumption that $\tau_{B} \ll \Delta t$ allows us to
neglect the mentioned correlations. As previously, we postpone the
discussion for {\em Auxiliary} sections. At present, we assume that
\begin{equation}
  \inro_{AB}(t) ~\approx~ \inro_{A}(t) \otimes \inro_{B}(t).
\label{me23} \end{equation}

By assumption, the reservoir (environment) is large,
its correlation time is very short, so the reservoir's relaxation
is fast. We may say that before any significant changes
occur in subsystem $\cal{A}$,  the reservoir would have enough time
to reach thermodynamic equilibrium. As it is known from statistical
physics such state is given as
\begin{equation}
   \sib
   = \sum_{z} p(z) \ket{z}\bra{z}
     \hspace*{10mm}\mathrm{where} \hspace*{10mm}
     p(z)
     = \frac{1}{\mathbb{Z}}
       \exp \left( - \frac{E_{z}}{k_{B}T} \: \right).
\label{me23b} \end{equation}
The quantity $Z$ is a partition sum
\begin{equation}
 \mathbb{Z}
 ~=~ \sum_{z} \exp \left( - \frac{E_{z}}{k_{B}T} \: \right).
\label{me23c} \end{equation}
States $\ket{z}$ and energies $E_{z}$ are the eigenstates
and eigenvalues of the reservoir hamiltonian, which
can be written as $H_{B} = \sum_{z} E_{z} \ket{z} \bra{z}$.
As a consequence we conclude that
\begin{equation}
   \bigl[ \: \sib, \; H_{B} \: \bigr] ~=~ 0,
\label{me23d} \end{equation}
so we can say that operator $\sib$ is stationary -- does not change
in time. It is worth noting that we could have reversed the
argument. First require stationarity, as expressed by \eqref{me23d}
which would entail relations \eqref{me23b} and \eqref{me23c}.
Moreover, commutation relation \eqref{me23d} implies
that the reduced density operator $\sib$ is of the same form
both in  Schr{\"{o}}dinger and interaction pictures.

Due to these remarks operator $\inro_{B}(t)$ appearing in
Eq.\eqref{me23} is simply replaced by  $\sib$. Therefore,
in Eq.\eqref{me16x} we make the replacement
$\inro_{AB} = \inro_{A}(t) \otimes \sib$.
So, we now have
\begin{align}
& \hspace*{-4mm}
  \Delta \inro_{A}(t)
   ~=~ \left( \frac{1}{i\hbar} \right)
       \int\limits_{t}^{t+\Delta t} dt_{1}
      ~\TTr_{B} \bigl[ \: \widetilde{V}_{AB}(t_{1}), \:
       \inro_{A}(t) \otimes \sib \:
       \bigr]
\nonumber \\
& \hspace*{1mm}
    + \left( \frac{1}{i\hbar} \right)^{2}
      \int\limits_{t}^{t+\Delta t} dt_{1}
      \int_{t}^{t_{1}} dt_{2} ~\TTr_{B}
      \bigl[ \: \widetilde{V}_{AB}(t_{1}), \;
      \bigl[ \: \widetilde{V}_{AB}(t_{2}), \;
      \inro_{A}(t) \otimes \sib \:
      \bigr].
\label{me27} \end{align}
The employed simplification facilitates computation of the
remaining traces. However, to proceed effectively,
we need to specify the interaction hamiltonian.

\section{Interaction hamiltonian and its properties}

\subsection{The form of $\widetilde{V}_{AB}(t)$}

Our next assumption concerns the shape of the interaction hamiltonian.
It is taken as (similarly as in the given references)
\begin{equation}
   V_{AB}
   ~=~ \sum_{\alpha} ~A_{\alpha} \otimes X_{\alpha}
   ~=~ \sum_{\alpha} ~A^{\dagger}_{\alpha}
       \otimes X^{\dagger}_{\alpha},
\label{me30} \end{equation}
where $A_{\alpha}$ are operators which act in the space of the states
of subsystem ${\cal A}$, while operators $X_{\alpha}$ correspond
to space of the reservoir's states. Operators appearing in the
definition \eqref{me30} need not be hermitian (each one separately).
Only the hamiltonian $V_{AB}$ must be hermitian. That is why we have
written the second equality. We can say that to each nonhermitian
term $A_{\alpha} \otimes X_{\alpha}$ corresponds the term
$A^{\dagger}_{\alpha} \otimes X^{\dagger}_{\alpha}$, and the latter
appears in the sum $V_{AB}$, but with another number. In {\em Auxiliary}
sections we will argue that it is not any limitation.
It is only important that the whole hamiltonian $V_{AB}$ must
be hermitian.

Operators $A_{\alpha}$ and $X_{\alpha}$ act in different spaces
so they are independent and commute.
In the interaction picture we immediately have
\begin{equation}
   \widetilde{V}_{AB}(t)
   ~=~\sum_{\alpha} ~\widetilde{A}_{\alpha}(t)
      \otimes \widetilde{X}_{\alpha}(t)
   ~=~\sum_{\alpha} ~\widetilde{A}^{\dagger}_{\alpha}(t)
      \otimes \widetilde{X}^{\dagger}_{\alpha}(t),
\label{me31} \end{equation}
with
\begin{equation}
  \widetilde{A}_{\alpha}(t)
  ~=~ e^{ iH_{A}t/\hbar} \; A_{\alpha} \; e^{-iH_{A}t/\hbar},
\hspace*{10mm}
  \widetilde{X}_{\alpha}(t)
  ~=~ e^{ iH_{B}t/\hbar} \; X_{\alpha} \; e^{-iH_{B}t/\hbar}.
\label{me32} \end{equation}
Rules of hermitian conjugation imply that
the conjugate operators transform to interaction picture in
the exactly the same manner as the initial ones.

We now make one more assumption about reservoir. We have already
taken $\inro_{B}(t) \approx \sib$. Here, we assume that in
the Schr{\"{o}}dinger picture
\begin{equation}
  \qav{X_{\alpha}}_{B}
  ~\equiv
  ~\TTr_{B} \left\{ \: X_{\alpha} \,\rho_{B}(t) \: \right\}
  ~=~ \TTr_{B} \left\{ \: X_{\alpha} \,\sib \: \right\}
  ~=~ 0.
\label{me37} \end{equation}
This assumption easily transforms to interaction picture
\begin{align}
  \qav{\widetilde{X}_{\alpha}(t)}_{B}
  &=\TTr_{B} \left\{ \:
     e^{ iH_{B}t/\hbar} \; X_{\alpha} \; e^{-iH_{B}t/\hbar}
     \; \sib \; \right\}
\nonumber \\
  &=\TTr_{B} \left\{ \: X_{\alpha}
        e^{-iH_{B}t/\hbar} \; \sib \; e^{iH_{B}t/\hbar} \;
        \right\}
  ~=~\TTr_{B} \left\{ \: X_{\alpha} \; \sib \; \right\}
  ~=~0,
\label{me39} \end{align}
which follows from commutation relation \eqref{me23d},
cyclic property of trace and \eqref{me37}.
This is rather a simplification, not a restrictive assumption.
This will be clarified and explained in {\em Auxiliary} sections.
Eq. \eqref{me37} (leading to \eqref{me39}) allows us to see that
the first term in the ME \eqref{me27} is, in fact, zero. Indeed
\begin{align}
& \hspace*{-4mm}
   \TTr_{B} \bigl[ \: \widetilde{V}_{AB}(t_{1}),
           ~\inro_{A}(t) \otimes \sib \: \bigr]
   = ~ \TTr_{B} \biggl[ \: \sum_{\alpha}
       \widetilde{A}_{\alpha}(t_{1})
       \otimes \widetilde{X}_{\alpha}(t_{1}),
       ~\inro_{A}(t) \otimes \sib \: \biggr]
\nonumber \\
& \hspace*{5mm}
    =~\sum_{\alpha}
      \Bigl\{ \: \widetilde{A}_{\alpha}(t_{1}) \inro_{A}(t) \:
      \TTr_{B} \Bigl[ \widetilde{X}_{\alpha}(t_{1}) \: \sib
            \Bigl]
\nonumber \\
& \hspace*{35mm}
      ~-~ \inro_{A} \widetilde{A}_{\alpha}(t_{1}) \:
      \TTr_{B} \Bigl[ \sib \: \widetilde{X}_{\alpha}(t_{1})
            \Bigl]
      \Bigl\} ~=~ 0.
\label{me40} \end{align}
Both traces are equal (cyclic property), nevertheless this
expression need not be zero, because  operators of the ${\cal A}$
system need not commute. If requirement \eqref{me37} is not fulfilled
then the above average may not  vanish. Assumption \eqref{me37}
and its consequence \eqref{me39} fortunately give zero, thus the
first term of Eq.\eqref{me27} vanishes and we remain with the
master equation
\begin{equation} \hspace*{-6mm}
   \Delta \, \inro_{A}(t)
    ~=~ \left( \: \frac{1}{i\hbar}\: \right)^{2}
        \int_{t}^{t+\Delta t} dt_{1}
        \int_{t}^{t_{1}} dt_{2}
       ~\TTr_{B} \left[ \: \widetilde{V}_{AB}(t_{1}), \;
        \bigl[ \: \widetilde{V}_{AB}(t_{2}),
       ~\inro_{A}(t) \otimes \sib\:
        \bigr] \right].
\label{me41a} \end{equation}
Expanding the commutators is simple. Moreover, one easily notices that
there are two pairs of hermitian conjugates. Hence we have
\begin{align}
& \hspace*{-4mm}
    \frac{\Delta \, \inro_{A}(t)}{\Delta t}
    ~=~ \frac{1}{\hbar^{2} \: \Delta t}
        \int\limits_{t}^{t+\Delta t} dt_{1}
        \int_{t}^{t_{1}} dt_{2}
       ~\TTr_{B} \; \Bigl\{
   \widetilde{V}_{AB}(t_{2}) \:
   \bigl( \inro_{A}(t) \otimes \sib \bigr) \:
   \widetilde{V}_{AB}(t_{1}) \:
\nonumber \\
& \hspace*{29mm}
 -~\widetilde{V}_{AB}(t_{1}) \:
   \widetilde{V}_{AB}(t_{2}) \:
   \bigl( \inro_{A}(t) \otimes \sib \bigr) \Bigr\}
   ~+~ \mathbb{H.C}.
\label{me41c} \end{align}
We can now use hamiltonian \eqref{me31} and perform further
transformations. It can be, however, shown \cite{al}--\cite{horn}
that this equation does not guarantee that the positivity of the
density operator $\inro_{A}(t)$ is preserved. It appears that
the so-called secular approximation is necessary. To perform
it effectively, it is convenient to present the interaction
hamiltonian in a somewhat different form.

\subsection{Operators $A_{\alpha}(\Omega)$}

Let us write the hamiltonian of the subsystem $\cal{A}$ as
\begin{equation}
    H_{A} ~=~ \sum_{a} \hbar \omega_{a} \ket{a}\bra{a}.
\label{me44} \end{equation}
States $\ket{a}$ constitute the complete and orthonormal basis in the
space of states of the subsystem~$\cal{A}$. The eigenfrequencies
$\omega_{a}$ may or may not be degenerate. We allow
$\omega_{a}=\omega_{b}$ for $a \neq b$. At present it suffices that
we distinguish different kets $\ket{a}$ solely by their
"{\em quantum number}" $a$. Similarly as in Refs.\cite{bp,horn},
we now define the operators $A_{\alpha}(\Omega)$ via the
following relation
\begin{equation}
   A_{\alpha}(\Omega)
   = \sum_{a,b} \delta(\omega_{ba} - \Omega) \:
     \ket{a} \elm{a}{A_{\alpha}}{b} \bra{b}.
\label{me45} \end{equation}
This representation may be called the decomposition of operator
$A_{\alpha}$ into eigenprojectors of hamiltonian $H_{A}$.
Delta $\delta(\omega_{ba} - \Omega)$ is of the Kronecker type,
that is
\begin{equation}
   \delta(\omega_{ba} - \Omega)
   =  \left\{
      \begin{array}{l}
      0 \quad \mathrm{for} \quad \omega_{ba} \ne \Omega, \\
      1 \quad \mathrm{for} \quad \omega_{ba} = \Omega,
      \end{array}  \right.
\label{me46} \end{equation}
In our considerations we allow for nonhermitian operators
$A_{\alpha}$. Hence, definition \eqref{me45} is augmented by the
following one
\begin{align}
   A^{\dagger}_{\alpha}(\Omega)
   &= \sum_{a,b} \delta(\omega_{ba} - \Omega) \:
     \ket{b} \elm{b}{A^{\dagger}_{\alpha}}{a} \bra{a}
\nonumber \\
   &= \sum_{a,b} \delta(\omega_{ab} - \Omega) \:
     \ket{a} \elm{a}{A^{\dagger}_{\alpha}}{b} \bra{b},
\label{me47} \end{align}
because it is always allowed to interchange the summation
indices $a \leftrightarrow b$. We stress that $A_{\alpha}(\Omega)$
contains Bohr frequency $\omega_{ba}$, while in
$A^{\dagger}_{\alpha}(\Omega)$ we have $\omega_{ab}=-\omega_{ba}$.
The following relation seems to be quite obvious
\begin{equation}
   \sum_{\Omega} \delta(\omega_{kn} - \Omega) =1,
\label{me48} \end{equation}
since out of all $\Omega$'s one will exactly match $\omega_{kn}$.
As a consequence we obtain
\begin{equation}
   \sum_{\Omega} A_{\alpha}(\Omega) = A_{\alpha}.
\label{me49} \end{equation}
Indeed, from definition \eqref{me45}, relation \eqref{me48}
and due to completeness of states $\ket{a}$ we get
\begin{align}
   \sum_{\Omega} A_{\alpha}(\Omega)
   &= \sum_{\Omega} \sum_{a,b}
       \delta(\omega_{ba} - \Omega) \:
       \ket{a} \elm{a}{A_{\alpha}}{b} \bra{b}
\nonumber \\
   &= \sum_{a,b} \ket{a} \elm{a}{A_{\alpha}}{b} \bra{b}
   =\idd A_{\alpha} \idd = A_{\alpha}.
\label{me50} \end{align}
Relation \eqref{me49} implies that the interaction hamiltonian
(in Schr{\"{o}}dinger picture) can be written as
\begin{equation}
   V_{AB}
   = \sum_{\alpha} A_{\alpha} \otimes X_{\alpha}
   = \sum_{\Omega} \sum_{\alpha}
     A_{\alpha}(\Omega) \otimes X_{\alpha}.
\label{me51} \end{equation}
Similarly as above we can also  show that
\begin{equation}
   \sum_{\Omega} A^{\dagger}_{\alpha}(\Omega)
   = A^{\dagger}_{\alpha},
\label{me52} \end{equation}
and
\begin{equation}
   V_{AB} = V^{\dagger}_{AB}
   = \sum_{\Omega} \sum_{\alpha} A^{\dagger}_{\alpha}(\Omega)
     \otimes X^{\dagger}_{\alpha}.
\label{me53} \end{equation}
Using definition \eqref{me45} we find the operator
$\widetilde{A}_{j}(\Omega)$ in the interaction picture
\begin{equation}
   \widetilde{A}_{\alpha}(\Omega)
   ~=~ e^{ iH_{A}t/\hbar} \; A_{\alpha}(\Omega) \;
       e^{-iH_{A}t/\hbar}
   ~=~ e^{ - i\Omega t} \; A_{\alpha}(\Omega),
\label{me55} \end{equation}
because $e^{ iH_{A}t/\hbar} \ket{a} = e^{ i\omega_{a}t} \ket{a}$.
Linking expressions \eqref{me51} and \eqref{me55} we write the
interaction hamiltonian in the interaction picture
\begin{align}
   \widetilde{V}_{AB}(t)
   &= \sum_{\Omega} \sum_{\alpha}
      e^{-i\Omega t} \: A_{\alpha}(\Omega) \otimes
      \widetilde{X}_{\alpha}(t)
\nonumber \\
   & = \sum_{\Omega} \sum_{\alpha}
     e^{i\Omega t} \: A^{\dagger}_{\alpha}(\Omega) \otimes
     \widetilde{X}^{\dagger}_{\alpha}(t)
   = \widetilde{V}^{\dagger}_{AB}(t).
\label{me58} \end{align}
Before starting to analyze ME \eqref{me41c}, let us notice that
operators $A_{\alpha}(\Omega)$ possess some interesting properties
\cite{bp}. Discussion of these properties is moved
to {\em Auxiliary} sections.

\subsection{Further analysis of master equation}

We return to master equation \eqref{me41c}. Interaction hamiltonian
$\widetilde{V}_{AB}(t_{2})$ is taken as in the first part
of Eq. \eqref{me58}, while  $\widetilde{V}_{AB}(t_{1})$ is
represented by its hermitian conjugate according to second
part of \eqref{me58}. This gives
\begin{align}
& \hspace*{-8mm}
   \frac{\Delta \, \inro_{A}(t)}{\Delta t}
   ~=~\frac{1}{\hbar^{2} \: \Delta t}
      \int\limits_{t}^{t+\Delta t} dt_{1}
      \int_{t}^{t_{1}} dt_{2} \sum_{\alpha,\beta}
      \sum_{\Omega,\Omega\,'}
     ~\TTr_{B} \; \Bigl\{
\nonumber \\
& \hspace*{14mm}
   e^{-i\Omega t_{2}}
   A_{\beta}(\Omega) \otimes \widetilde{X}_{\beta}(t_{2})
   \bigl[ \inro_{A}(t) \otimes \sib \bigr]
   e^{i\Omega\,' t_{1}}
   A^{\dagger}_{\alpha}(\Omega\,') \otimes
   \widetilde{X}^{\dagger}_{\alpha}(t_{1})
\nonumber \\[2mm]
& \hspace*{-4mm}
   -~e^{i\Omega\,' t_{1}}
   A^{\dagger}_{\alpha}(\Omega\,') \otimes
   \widetilde{X}^{\dagger}_{\alpha}(t_{1})
   \bigl[ e^{-i\Omega t_{2}} A_{\beta}(\Omega)
          \otimes \widetilde{X}_{\beta}(t_{2}) \bigr]
   \inro_{A}(t) \otimes \sib
   \Bigr\} ~+~ \mathbb{H.C}.
\label{me60} \end{align}
Computing tensor products we remember that partial trace is
taken only with respect to reservoir variables. Moreover,
we note that these traces are the same (cyclic property).
Therefore we denote
\begin{equation}
  \bar{G}_{\alpha \beta}(t_{1} - t_{2} )
  ~=~ \TTr_{B} \left\{ \: \widetilde{X}^{\dagger}_{\alpha}(t_{1})
      \: \widetilde{X}_{\beta}(t_{2}) \: \sib  \: \right\}.
\label{me62} \end{equation}
Finally we rewrite the arguments of the exponentials as
$i\Omega\,' t_{1} - i \Omega t_{2} =%
i(\Omega\,' - \Omega) t_{1} + i \Omega (t_{1} - t_{2})$.
Thus Eq.\eqref{me60} becomes
\begin{align}
& \hspace*{-4mm}
   \frac{\Delta \, \inro_{A}(t)}{\Delta t}
  ~=~ \frac{1}{\hbar^{2} \: \Delta t}
      \int\limits_{t}^{t+\Delta t} dt_{1}
      \int_{t}^{t_{1}} dt_{2}
      \sum_{\alpha,\beta} \sum_{\Omega,\Omega\,'}
      ~e^{i(\Omega\,' - \Omega) t_{1}}
      ~e^{ i \Omega (t_{1} - t_{2})}
      \bar{G}_{\alpha \beta}(t_{1} - t_{2})
\nonumber \\
& \hspace*{19mm} \times
   \Bigl[ \: A_{\beta}(\Omega) \inro_{A}(t)
             A^{\dagger}_{\alpha}(\Omega\,')
         ~-~ A^{\dagger}_{\alpha}(\Omega\,')
             A_{\beta}(\Omega) \inro_{A}(t)
   \Bigr] ~+~ \mathbb{H.C}.
\label{me64} \end{align}
The quantity $\bar{G}_{\alpha \beta}(t_{1}-t_{2})$
is called the correlation function of the reservoir.
We will briefly discuss its properties.

\subsection{Correlation functions  $\bar{G}_{\alpha \beta}$}

Let us focus on the functions defined by the right
hand side of Eq.\eqref{me62}, that is
\begin{equation}
   G_{\alpha \beta}(t_{1},t_{2})
   ~=~ \TTr_{B} \left\{ \widetilde{X}^{\dagger}_{\alpha}(t_{1}) \:
       \widetilde{X}_{\beta}(t_{2}) \: \sib \right\}.
\label{me65a} \end{equation}
These are the functions of two variables and it is not,
{\em a priori}, clear that they depend only of the difference
$\tau=t_{1}-t_{2}$. Before discussing this fact,
let us note that $G^{\ast}_{\alpha \beta}(t_{1},t_{2})%
~=~ G_{\beta \alpha}(t_{2},t_{1})$. To prove this relation,
we recall that $\TTr_{B}^{\ast}\{A\}=\TTr_{B}\{A^{\dagger}\}$,
so that the definition \eqref{me65a} gives
\begin{align}
   G^{\ast}_{\alpha \beta}(t_{1},t_{2})
   &~=~ \TTr_{B} \left\{ \sib \widetilde{X}^{\dagger}_{\beta}(t_{2})
       \: \widetilde{X}_{\alpha}(t_{1}) \: \right\}
\nonumber \\
   &~=~ \TTr_{B} \left\{ \widetilde{X}^{\dagger}_{\beta}(t_{2})
       \: \widetilde{X}_{\alpha}(t_{1}) \: \sib \right\}
   ~=~ G_{\beta \alpha}(t_{2},t_{1}).
\label{me65c} \end{align}

Now, we will show that the function $G_{\alpha\beta}(t_{1},t_{2})$
is indeed a function of the difference of its arguments.
The key role plays the fact that the state of the reservoir
(density operator $\sib$) is stationary (does not change in time).
Explicitly using the interaction picture we get
\begin{equation} \hspace*{-4mm}
  G_{\alpha \beta}(t_{1},t_{2})
  = \TTr_{B} \left\{ \,
        \bigl( e^{ iH_{B}t_{1}/\hbar} \: X^{\dagger}_{\alpha} \:
               e^{-iH_{B}t_{1}/\hbar} \bigr)
       \:\bigl( e^{ iH_{B}t_{2}/\hbar} \: X_{\beta} \:
               e^{-iH_{B}t_{2}/\hbar} \bigr)  \:
               \sib \: \right\}.
\label{me65d} \end{equation}
The trace is cyclic and $\sib$ commutes with hamiltonianem $H_{B}$
so we conclude that
\begin{align}
  G_{\alpha\beta}(t_{1},t_{2})
  &= \TTr_{B} \left\{ \:
        e^{ iH_{B}(t_{1}-t_{2})/\hbar} \; X^{\dagger}_{\alpha} \;
        e^{-iH_{B}(t_{1}-t_{2})/\hbar} \; X_{\beta} \:
        \sib \: \right\}
\nonumber \\
  &=\TTr_{B} \left\{ \:
        \widetilde{X}^{\dagger}_{\alpha}(t_{1}-t_{2}) \,
        \widetilde{X}_{\beta}(0)
        \: \sib \: \right\}
  ~=~ \bar{G}_{\alpha \beta}(\tau = t_{1} - t_{2}),
\label{me65e} \end{align}
for two moments of time $t_{1} > t_{2}$. Reservoir's correlation
function effectively depends only on one variable. This fact is
denoted by a bar over the symbol of correlation function.
Thus we write
\begin{equation}
  G_{\alpha \beta}(t_{1},t_{2})
  ~=~ \bar{G}_{\alpha \beta}(\tau)
  ~=~ \TTr_{B} \left\{ \:
      \widetilde{X}_{\alpha}^{\dagger}(\tau) \,
                 X_{\beta} \: \sib \: \right\}
\label{me65f} \end{equation}
Such correlation functions are called stationary.
Sometimes the concept of stationarity means invariance
with respect to time translation. Indeed, for arbitrary
time $T$ we have
\begin{equation}
  \bar{G}_{\alpha \beta}(t_{1}+T,t_{2}+T)
  ~=~ \bar{G}_{\alpha \beta} \bigl( (t_{1}+T) - (t_{2}+T) \bigr)
  ~=~ \bar{G}_{\alpha \beta}(t_{1} - t_{2}).
\label{me65g} \end{equation}
This property of the correlation functions is a straightforward
consequence of the stationarity of reservoir's density
operator $\sib$.

\section{Discussion of times}

Preparing this  section of our tutorial we greatly benefited
from the analogous discussion by Cohen-Tannoudji {\em et al}
\cite{coh}. To large extent we follow their reasoning,
trying to elucidate some less obvious points.
Admitting this, we will refrain from giving multiple
references to their work.

\subsection{Limits of the integrals and Markov approximation}

\begin{figure}[ht]
\begin{center}
\scalebox{0.6}[0.6]{\includegraphics{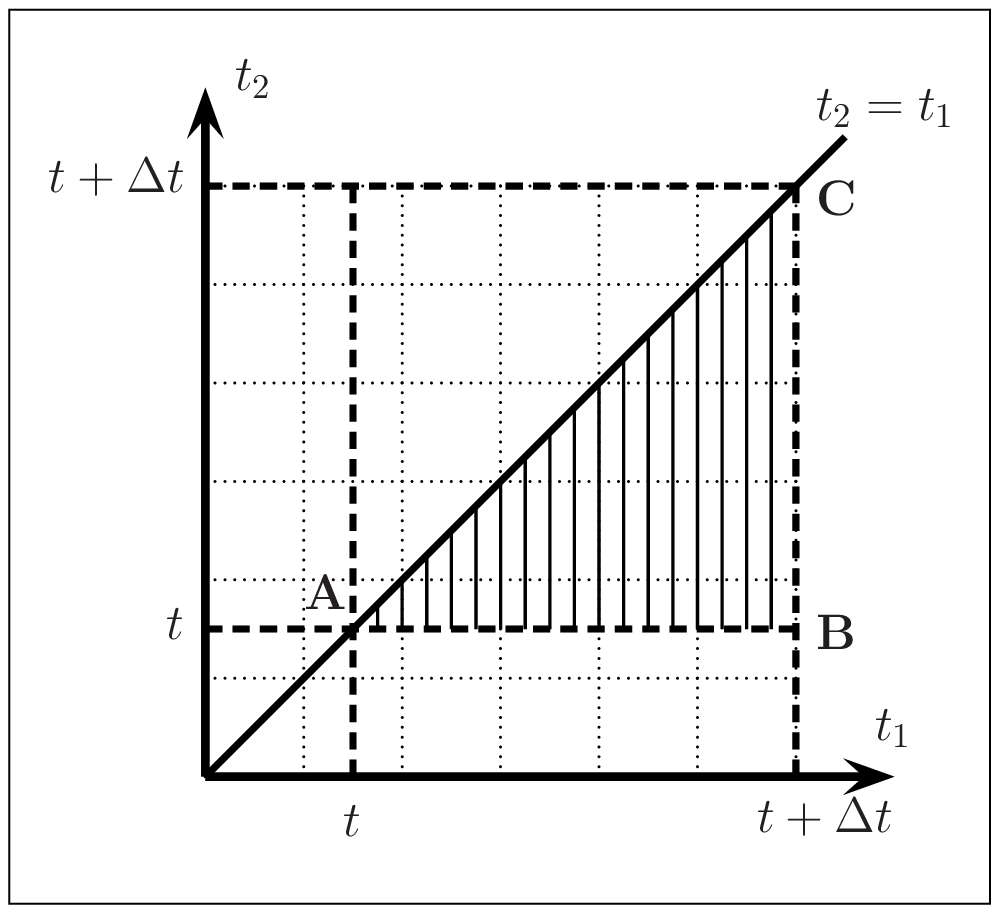}}
\scalebox{0.6}[0.6]{\includegraphics{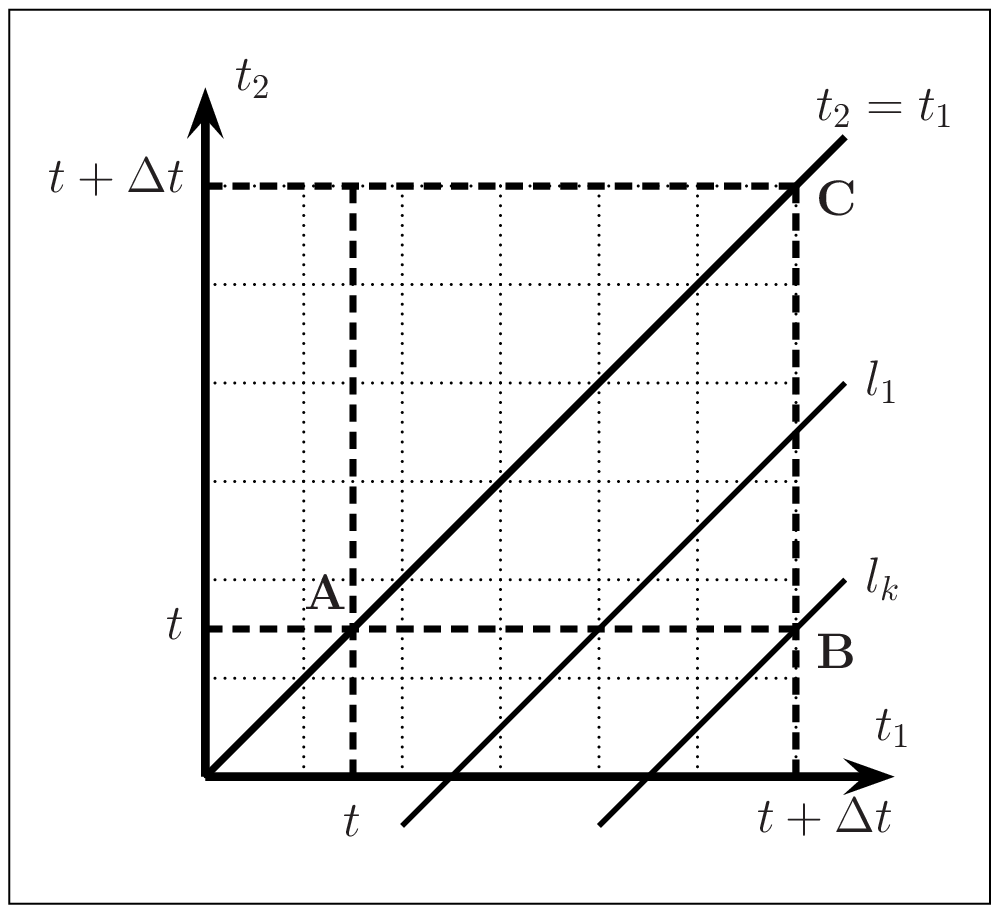}}
\parbox{100mm}{\setlength{\baselineskip}{\skbaseline}
\caption{\small Left figure presents the integration region
in the double integral in Eq.\eqref{me64}. Right figure
illustrates the change to new variables $\tau=t_{1}-t_{2}$
and $t_{1}$. Other explanations are to be found in the text.}
\label{xmerys01}
}
\end{center}
\end{figure}

In master equation \eqref{me64} one integrates over the triangle
$ABC$ which is shown in Fig.~\ref{xmerys01}. Firstly, one computes
the integral over $dt_{2}$ in the range from $t$ to $t_{1}$.
This is indicated by thin vertical lines (at left).
Next, one sums such contributions by integrating over $dt_{1}$
from $t$ to $t+\Delta t$.
The integrand in \eqref{me64} contains correlation functions of the
reservoir which depend on the difference $\tau=t_{1}-t_{2}$.
We stress that we always have $t_{1} \ge t_{2}$, so that
$\tau \ge 0$. The integration over the triangle can be performed in
another manner.

Let us consider the geometry (right graph in Fig. \ref{xmerys01}).
Along the diagonal AC
we have $t_{1}=t_{2}$, so $\tau=t_{1}-t_{2}=0$. The straight line
$l_{1}$ has  (in $t_{1}$ and $t_{2}$ variables) the equation
$t_{2}=t_{1} - \tau$, where $\tau$ is fixed, since $(-\tau)$
is the coordinate $t_{2}$ of the point where the discussed line
intersects the axis $t_{2}$. Then, for the line $l_{k}$
(passing through point B) $\tau$ is also fixed (by the same
argument, as in the case of line $l_{1}$). On $l_{k}$, at the
point B we have  $t_{1}=t+\Delta t$ and $t_{2}=t$.
Thus, we have $\tau=\Delta t$.
Parametr $\tau$ specifies the skew straight lines (parallel to the
diagonal AC) and passing through triangle ABC.
\begin{figure}[ht]
\begin{center}
\scalebox{0.6}[0.6]{\includegraphics{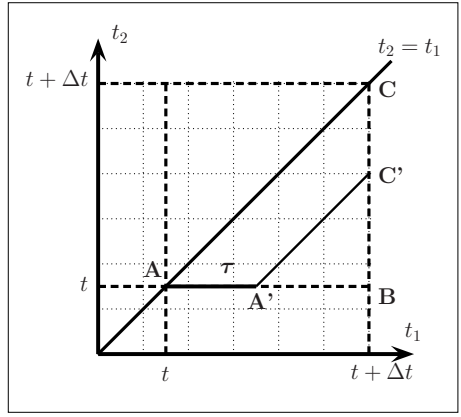}}
\parbox{100mm}{\setlength{\baselineskip}{\skbaseline}
\caption{\small Illustration of the change of integration variables
in equation \eqref{me64}, Transformation to new variables
$\tau=t_{1}-t_{2}$ i $t_{1}$. Other explanations -- in the text.}
\label{xmerys02}
}
\end{center}
\end{figure}
Integration over the triangle ABC is now done as follows.
We fix $\tau \in (0, \Delta t)$ and we move along the segment A'C'
(see Fig.\ref{xmerys02}). Variable $t_{1}$ runs in the interval
from $t+\tau$ to $t+\Delta t$. So, we first integrate over $dt_{1}$
from $t+\tau$ to $t+\Delta t$ (along the segment A'C').
Next, we integrate over $d\tau$ from zero to $\Delta t$.
In this manner we sum the contributions from all skew segments
covering the triangle ABC. Therefore, we can write
\begin{equation}
   \int\limits_{t}^{t+\Delta t} dt_{1}
   \int_{t}^{t_{1}} dt_{2}
   ~=~\int_{0}^{\Delta t} d\tau
      \int\limits_{t+\tau}^{t+\Delta t} dt_{1},
\label{me70} \end{equation}
while we remember that $\tau= t_{1}-t_{2}$ (or $t_{2}=t_{1}-\tau$).
Performing the discussed changes of integration variables in
Eq. \eqref{me64}, we get
\begin{align}
& \hspace*{-4mm}
   \frac{\Delta \, \inro_{A}(t)}{\Delta t}
  ~=~ \frac{1}{\hbar^{2} \: \Delta t}
      \int\limits_{0}^{\Delta t} d\tau
      \int\limits_{t+\tau}^{t+\Delta t} dt_{1}
      \sum_{\alpha,\beta} \sum_{\Omega,\Omega\,'}
      ~e^{i(\Omega\,' - \Omega) t_{1}}
      ~e^{ i \Omega \tau}
      \bar{G}_{\alpha \beta}(\tau)
\nonumber \\
& \hspace*{12mm} \times
   \Bigl[ \: A_{\beta}(\Omega) \inro_{A}(t)
             A^{\dagger}_{\alpha}(\Omega\,')
         ~-~ A^{\dagger}_{\alpha}(\Omega\,')
             A_{\beta}(\Omega) \inro_{A}(t)
   \Bigr] ~+~ \mathbb{H.C}.,
\label{me71} \end{align}
We recall that the considered time intervals satisfy
the requirement $\Delta t \gg \tau_{B}$ (which will be discussed in
detail later). If it is true, then the main contribution to the
integral over $d\tau$ in Eq.\eqref{me71} will come from the region
in the neighborhood of $0 \le \tau < \tau_{B} \ll \Delta t$.
Geometrically, this corresponds to a narrow belt which is parallel
to the diagonal AC and lies just below it. It follows,
that outside this region the reservoir's correlation functions
practically vanish (decay to zero). Therefore, we will not make any
serious error moving the upper limit of integration over $d\tau$
to infinity. Moreover, since only small $\tau$'s contribute
significantly, the lower limit of the integral over $dt_{1}$
may be approximated simply by $t$, so only a small "initial"
error might be introduced. With these approximations equation
\eqref{me71} yields
\begin{align}
& \hspace*{-4mm}
   \frac{\Delta \, \inro_{A}(t)}{\Delta t}
  ~=~ \frac{1}{\hbar^{2} \: \Delta t}
      \sum_{\Omega,\Omega\,'} \sum_{\alpha,\beta}
      \int_{0}^{\infty} d\tau
           ~e^{ i \Omega \tau} \bar{G}_{\alpha \beta}(\tau)
      \int\limits_{t}^{t+\Delta t} dt_{1}
           ~e^{i(\Omega\,' - \Omega) t_{1}}
\nonumber \\
& \hspace*{12mm} \times
   \Bigl[ \: A_{\beta}(\Omega) \inro_{A}(t)
             A^{\dagger}_{\alpha}(\Omega\,')
         ~-~ A^{\dagger}_{\alpha}(\Omega\,')
             A_{\beta}(\Omega) \inro_{A}(t)
   \Bigr] ~+~ \mathbb{H.C}.
\label{me72} \end{align}
Introducing the quantities
\begin{subequations} \label{me73}
\begin{align}
  J(\Omega\,' - \Omega)
  &= \frac{1}{\Delta t} \;
      \int\limits_{t}^{t + \Delta t}  dt_{1}
        ~\exp\bigl[ \: i(\Omega\,' - \Omega) t_{1} \: \bigl],
\label{me73a} \\
  W_{\alpha\beta}(\Omega)
  &= \int_{0}^{\infty} d\tau
     ~e^{i \Omega \tau} ~\bar{G}_{\alpha \beta}(\tau)
 = \int_{0}^{\infty} d\tau
     ~e^{i \Omega \tau}
     \: \TTr_{B} \bigl\{ \widetilde{X}^{\dagger}_{\alpha}(\tau) \,
     X_{\beta} \sib \bigr\},
\label{me73b}
\end{align} \end{subequations}
we rewrite Eq.\eqref{me72} as follows
\begin{align}
& \hspace*{-4mm}
   \frac{\Delta \, \inro_{A}(t)}{\Delta t}
  ~=~ \frac{1}{\hbar^{2}}
      \sum_{\Omega,\Omega\,'} \sum_{\alpha,\beta}
      J(\Omega\,'- \Omega) W_{\alpha\beta}(\Omega)
\nonumber \\
& \hspace*{12mm} \times
   \Bigl[ \: A_{\beta}(\Omega) \inro_{A}(t)
             A^{\dagger}_{\alpha}(\Omega\,')
         ~-~ A^{\dagger}_{\alpha}(\Omega\,')
             A_{\beta}(\Omega) \inro_{A}(t)
   \Bigr] ~+~ \mathbb{H.C}.
\label{me74} \end{align}
This equation specifies the rate of change of the reduced density
operator $\inro_{A}(t)$ within the time interval $(t,t+\Delta t)$.
The quotient $\Delta \inro_{A}(t)/\Delta t$ can be treated as an
averaging
\begin{equation}
   \frac{\Delta \, \inro_{A}(t)}{\Delta t}
   ~=~\frac{\inro_{A}(t+\Delta t) - \inro_{A}(t)}{\Delta t}
   ~=~ \frac{1}{\Delta t}
       \int_{t}^{t+\Delta t} dt_{1}
       ~\frac{d\, \inro_{A}(t_{1})}{dt_{1}}.
\label{me75} \end{equation}
This averaging results in smoothing all very rapid changes of
$\inro_{A}(t)$ which may occur during the interval $(t,\;t+\Delta t)$.
In principle we should account for such rapid changes.
We do not do that because right hand side of Eq.\eqref{me74}
contains $\inro_{A}(t)$, while the left hand side represents
the smoothed rate of change. This rate depends on the density
operator $\inro_{A}$ in the past, that is at the moment when the
smoothed evolution was started. So our next approximation
consists in replacing the smoothed rate by a usual derivative.
In other words, the variation at an instant $t$ (that is
the derivative $d\inro_{A}(t)/dt$) is connected with the
value of $\inro_{A}(t)$ at the very same instant. This
approximation allows us to use a usual derivative at
the left hand side of \eqref{me74}. This approximation sometimes
is called a Markovian one since it connects the variations
of some physical quantity with its value at the same instant,
independently from the values which this quantity had at earlier
moments. We can say that Markovian approximation consists in
neglecting the influence of the history of the physical system
on its current state which fully determines the presently
occurring changes. In some literature sources this approximation
is also called the coarse-graining one, because small and rapid
fluctuations are neglected when the evolution is investigated
on a much longer time scale specified by $\Delta t$.

With all the discussed approximation our master equation
\eqref{me74} becomes
\begin{align}
& \hspace*{-4mm}
   \ddt \, \inro_{A}(t)
  ~=~ \frac{1}{\hbar^{2}}
      \sum_{\Omega,\Omega\,'} \sum_{\alpha,\beta}
      J(\Omega\,'- \Omega) W_{\alpha\beta}(\Omega)
\nonumber \\
& \hspace*{12mm} \times
   \Bigl[ \: A_{\beta}(\Omega) \inro_{A}(t)
             A^{\dagger}_{\alpha}(\Omega\,')
         ~-~ A^{\dagger}_{\alpha}(\Omega\,')
             A_{\beta}(\Omega) \inro_{A}(t)
   \Bigr] ~+~ \mathbb{H.C}.
\label{me76} \end{align}

\subsection{Schr{\"{o}}dinger picture}

At this stage, we return  to the Schr{\"{o}}dinger picture and we
insert
\begin{equation}
  \inro_{A}(t)
  ~=~ e^{ iH_{A}t/\hbar} \; \rho_{A}(t) \;
      e^{-iH_{A}t/\hbar},
\label{me77} \end{equation}
into Eq. \eqref{me76}.
When computing the derivative at the left hand side we reproduce
the free evolution term. Thus, from \eqref{me76} we get
\begin{align}
& \hspace*{-1mm}
    e^{iH_{A}t/\hbar} \left( \ddt \rho_{A}(t) \right)
    e^{-iH_{A}t/\hbar}
    ~=~ -\: \frac{i}{\hbar} \; e^{iH_{A}t/\hbar}
         \bigl[ \: H_{A}, ~\rho_{A}(t) \: \bigr]
         e^{-iH_{A}t/\hbar}
\nonumber \\
& \hspace*{1mm}
   +~\biggr\{ \;
     \frac{1}{\hbar^{2}}
     \sum_{\Omega,\Omega\,'} \sum_{\alpha,\beta}
     J(\Omega\,'- \Omega) W_{\alpha\beta}(\Omega)
     \Bigl[ \: A_{\beta}(\Omega) \:
               e^{iH_{A}t/\hbar}\:\rho_{A}(t)\:
                   e^{-iH_{A}t/\hbar}
               A^{\dagger}_{\alpha}(\Omega\,')
\nonumber \\
& \hspace*{29mm}
   -~ A^{\dagger}_{\alpha}(\Omega\,') A_{\beta}(\Omega)
      e^{iH_{A}t/\hbar} \: \rho_{A}(t) \: e^{-iH_{A}t/\hbar}
      \Bigr] ~+~ \mathbb{H.C} \; \biggr\}.
\label{me78} \end{align}
Multiplying on the left by $e^{-iH_{A}t/\hbar}$ and on the right
by $e^{iH_{A}t/\hbar}$, we use relation \eqref{me55} and
its hermitian conjugate (for negative times). This yields
\begin{align}
& \hspace*{-1mm}
    \ddt \rho_{A}(t)
   ~= -\: \frac{i}{\hbar} \bigl[ \: H_{A}, ~\rho_{A}(t) \: \bigr]
      +~\biggr\{ \;
        \frac{1}{\hbar^{2}}
        \sum_{\Omega,\Omega\,'} \sum_{\alpha,\beta}
        J(\Omega\,'- \Omega) W_{\alpha\beta}(\Omega)
        ~e^{i(\Omega - \Omega\,')t}
\nonumber \\
& \hspace*{12mm} \times
   \Bigl[ \: A_{\beta}(\Omega) \rho_{A}(t)
             A^{\dagger}_{\alpha}(\Omega\,')
          ~-~A^{\dagger}_{\alpha}(\Omega\,')
             A_{\beta}(\Omega) \rho_{A}(t)
   \: \Bigr] ~+~ \mathbb{H.C} \; \biggr\}.
\label{me81} \end{align}

\subsection{Integral $J(\Omega\,'-\Omega)$ and secular approximation}

Our master equation contains the integral $J(\Omega\,'-\Omega)$
defined in \eqref{me73a}. Its computation is straightforward.
Denoting temporarily $x=\Omega\,'-\Omega$, we get
\begin{equation}
  J(x)
  = \int_{t}^{t+\Delta t} dt_{1}
     ~\frac{~e^{ixt_{1}}~}{\Delta t}
  = e^{ixt+ix\Delta t/2} \;
      \frac{\sin\left( \frac{x\Delta t}{2}\right)}
           {\left( \frac{x\Delta t}{2} \right)}
  = e^{ixt} \; F(x),
\label{me84} \end{equation}
where we have introduced (as in \cite{coh}) a function specified as
\begin{equation}
   F(x) ~=~ e^{ix\Delta t/2}~
            \frac{\sin\left( \frac{x\Delta t}{2}\right)}
                 {\left( \frac{x\Delta t}{2} \right)}.
\label{me85} \end{equation}
Due to the obtained results we can write
\begin{equation}
   J(\Omega\,' - \Omega)
  ~=~ e^{i(\Omega\,' - \Omega)t} \:
      F(\Omega\,' - \Omega).
\label{me86} \end{equation}
Inserting the computed integral into\eqref{me81} we note
that the exponential factor cancels out. Hence
\begin{align}
& \hspace*{-1mm}
  \ddt \, \rho_{A}(t)
  ~=~ -\:\frac{i}{\hbar} \:
       \bigl[ \: H_{A}, ~\rho_{A}(t) \: \bigr]
     + \biggr\{ \: \frac{1}{\hbar^{2}}
       \sum_{\Omega\,',\Omega} \sum_{\alpha,\beta}
       F(\Omega\,'-\Omega) W_{\alpha\beta}(\Omega)
\nonumber \\
& \hspace*{12mm} \times
    \Bigl[ \: A_{\beta}(\Omega) \: \rho_{A}(t) \:
              A^{\dagger}_{\alpha}(\Omega)
           ~-~A^{\dagger}_{\alpha}(\Omega) \:
              A_{\beta}(\Omega) \: \rho_{A}(t)  \: \Bigr]
   ~+~ \mathbb{H.C} \: \biggr\}.
\label{me88}
\end{align}
The sense of function $F(\Omega\,'-\Omega)$ must be carefully
considered. We proceed along the lines similar to those in \cite{coh}.
It is easy to see that function $|F(x)|$ has a sharp
maximum for $x=\Omega\,' - \Omega=0$, where it is equal to unity.
\begin{figure}[ht]
\begin{center}
\scalebox{0.6}[0.6]{\includegraphics{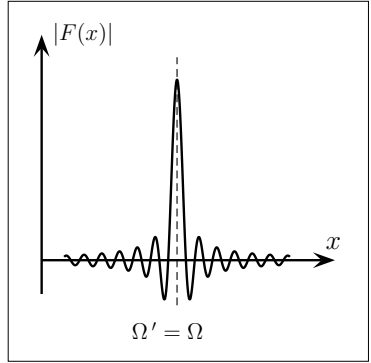}}
\parbox{130mm}{\setlength{\baselineskip}{\skbaseline}
\caption{\small The graph of the modulus of the function
$F(\Omega\,'-\Omega)$ which appears in \eqref{me88}.
If the time  $\Delta t$ is sufficiently large then the graph
has a sharp maximum for $\Omega\,'=\Omega$.}
\label{xmerys03}
}
\end{center}
\end{figure}

\noindent
Zeroes of this function correspond to
\begin{equation}
   \pol x \Delta t = n \pi
   \hspace*{6mm} \Longrightarrow \hspace*{6mm}
   x = \frac{2n\pi}{\Delta t}.
\label{me89} \end{equation}
If the time $\Delta t$ is sufficiently long then the central
maximum is very narrow. The question is what does it mean
"sufficiently long time". Let us consider two possibilities.
\begin{enumerate} \parskip=-1mm
\item If $x = \bigl| \Omega\,' - \Omega \bigr| \ll %
      (\Delta t)^{-1}$, the argument of function $|F(x)|$ is very
      close to zero, its value being practically one.
\item If $x = \bigl| \Omega\,' - \Omega \bigr| \gg %
      (\Delta t)^{-1}$ (Bohr frequencies are significantly different)
      then $|F(x)|$ is close to zero, as it is seen in
      Fig \ref{xmerys03}.
\end{enumerate}
We conclude that the terms at the right hand side of master
equation \eqref{me88} containing  the operator products
$A^{\dagger}_{\alpha}(\Omega\,')\,A_{\beta}(\Omega)$, for which
$\bigl| \Omega\,' - \Omega \bigr| \gg (\Delta t)^{-1}$
practically do not contribute to the evolution of the density
operator $\rho_{A}(t)$.
According to the first possibility above, significant contributions
come only from such couplings that operators
$A_{\alpha}^{\dagger}(\Omega\,')$
and $A_{\beta}(\Omega)$ have practically equal Bohr frequencies.

As we know, time $T_{A}$ is a characteristic relaxation time
in subsystem ${\cal A}$ due to interaction with reservoir
and it satisfies the estimate $\Delta t \ll T_{A}$
(we discuss it later). It can be argued (see also \cite{coh})
that the terms in
master equation \eqref{me88}, in which $|\Omega\,' - \Omega|%
\sim (\Delta t)^{-1}$ also give very small contributions,
so that they can be neglected. As a result of all these
approximations, we may say that only those terms in right hand side
of master equation \eqref{me88} contribute significantly
for which $\bigl| \Omega\,' - \Omega \bigr| = 0$. Such an
approximation is called the secular one. It allows us to
replace the function $F(\Omega\,' - \Omega)$ by the Kronecker
delta $\delta(\Omega\,' - \Omega)$ defined as in \eqref{me46}.
It reminds us that  only the terms satisfying the requirement
$(\Omega\,' = \Omega)$ give nonzero contribution.
All these arguments lead to master equation of the form
\begin{align}
& \hspace*{-8mm}
   \ddt \, \rho_{A}(t)
  ~=~ -\:\frac{i}{\hbar} \:
       \bigl[ \: H_{A}, ~\rho_{A}(t) \: \bigr]
    ~+~\biggr\{ \: \frac{1}{\hbar^{2}}
       \sum_{\alpha,\beta} \sum_{\Omega\,',\Omega}
       \delta(\Omega\,'-\Omega)  \: W_{\alpha\beta}(\Omega)
\nonumber \\
& \hspace*{12mm} \times
   \Bigl[ \: A_{\beta}(\Omega) \: \rho_{A}(t) \:
             A^{\dagger}_{\alpha}(\Omega\,')
          ~-~A^{\dagger}_{\alpha}(\Omega\,') \:
             A_{\beta}(\Omega) \: \rho_{A}(t)  \: \Bigr]
  ~+~ \mathbb{H.C} \: \biggr\}.
\label{me90} \end{align}
The presence of the discussed Kronecker delta simplifies
one of the summations, which gives
\begin{align}
& \hspace*{-8mm}
\ddt \, \rho_{A}(t)
  ~= -\:\frac{i}{\hbar}
       \bigl[ H_{A}, \: \rho_{A}(t) \bigr]
   +~\biggr\{ \:
     \frac{1}{\hbar^{2}}
     \sum_{\alpha,\beta} \sum_{\Omega} W_{\alpha\beta}(\Omega)
     \Bigl[ A_{\beta}(\Omega) \rho_{A}(t)
            A^{\dagger}_{\alpha}(\Omega)
\nonumber \\
& \hspace*{55mm}
        ~-~ A^{\dagger}_{\alpha}(\Omega)
            A_{\beta}(\Omega) \rho_{A}(t)
     \Bigr] + \mathbb{H.C} \;
     \biggr\}.
\label{me91} \end{align}
The fundamental part of the microscopic derivation of the master
equation is finished. We shall perform some transformations
which have important, but rather cosmetic character.
We want to transform master equation into the so-called standard
form. All other comments are, as mentioned many times,
left to {\em Auxiliary} sections.

\section{Standard form}

\subsection{Introduction}

Standard form is important, because it can be shown
(see \cite{al,bp,horn}) that this form
guarantees preservation of hermiticity, normalization and, first
of all, the positivity of the reduced density operator $\rho_{A}$.
If our master equation \eqref{me91} can be brought into the
standard form, then we can be sure that all the necessary
properties of the density operator of subsystem $\cal{A}$
are indeed preserved. Obviously, the first term in the right
hand side of equation \eqref{me91} describes the unitary
evolution, hence we shall concentrate only on the second term.
Writing explicitly the hermitian conjugates, we have
\begin{align}
& \hspace*{-8mm}
  \ddt \, \rho_{A}(t) \Bigl |_{d.}
  =\frac{1}{\hbar^{2}}
     \sum_{\Omega} \sum_{\alpha,\beta}
     W_{\alpha\beta}(\Omega)
     \Bigl[ \: A_{\beta}(\Omega) \rho_{A}(t)
               A^{\dagger}_{\alpha}(\Omega)
            ~-~A^{\dagger}_{\alpha}(\Omega)
               A_{\beta}(\Omega) \rho_{A}(t)
     \: \Bigr]
\nonumber \\
 &+ \frac{1}{\hbar^{2}}
     \sum_{\Omega} \sum_{\alpha,\beta}
     W^{\ast}_{\alpha\beta}(\Omega)
     \Bigl[ \: A_{\alpha}(\Omega) \rho_{A}(t)
               A^{\dagger}_{\beta}(\Omega)
            ~-~\rho_{A}(t) A^{\dagger}_{\beta}(\Omega)
               A_{\alpha}(\Omega)
     \: \Bigr],
\label{me96a} \end{align}
because operator $\rho_{A}(t)$ is hermitian (the proof that
hermiticity is preserved will be presented in {\em Auxiliary}
sections). In the second term we interchange the summation
indices $\alpha \leftrightarrow \beta$ which gives
\begin{align}
& \hspace*{-8mm}
  \ddt \, \rho_{A}(t) \Bigl |_{d.}
  =\frac{1}{\hbar^{2}}
     \sum_{\Omega} \sum_{\alpha,\beta}
     W_{\alpha\beta}(\Omega)
     \Bigl[ \: A_{\beta}(\Omega) \rho_{A}(t)
               A^{\dagger}_{\alpha}(\Omega)
            ~-~A^{\dagger}_{\alpha}(\Omega)
               A_{\beta}(\Omega) \rho_{A}(t)
     \: \Bigr]
\nonumber \\
& \hspace*{4mm}
   + \frac{1}{\hbar^{2}}
     \sum_{\Omega} \sum_{\alpha,\beta}
     W^{\ast}_{\beta \alpha}(\Omega)
     \Bigl[ \: A_{\beta}(\Omega) \rho_{A}(t)
               A^{\dagger}_{\alpha}(\Omega)
            ~-~\rho_{A}(t) A^{\dagger}_{\alpha}(\Omega)
               A_{\beta}(\Omega)
     \: \Bigr].
\label{me96b} \end{align}

\subsection{New notation}

For further convenience we introduce the following notation
\begin{subequations} \label{me97}
\begin{align}
    \Gamma_{\alpha \beta}(\Omega)
    &= W_{\alpha\beta}(\Omega)
    ~+~ W^{\ast}_{\beta\alpha}(\Omega),
\label{me97a} \\
    \Delta_{\alpha \beta}(\Omega)
    &= \frac{1}{2i} \bigl[ W_{\alpha\beta}(\Omega)
    ~-~ W^{\ast}_{\beta\alpha}(\Omega) \bigr].
\label{me97b}
\end{align} \end{subequations}
The matrix $\Gamma_{\alpha \beta}(\Omega)$ is hermitian and
positively defined. The latter property is difficult to prove.
It requires some advanced mathematics and we take this
fact for granted. The readers are referred to literature
\cite{al,bp}. Hermiticity of $\Gamma_{\alpha \beta}$
follows directly from the definition \eqref{me97a}.
Indeed, we have
\begin{equation}
   \Gamma^{\ast}_{\alpha \beta}(\Omega)
   ~=~W^{\ast}_{\alpha \beta}(\Omega)
      ~+~ W_{\beta \alpha}(\Omega)
   ~=~ \Gamma_{\beta \alpha}(\Omega).
\label{me98} \end{equation}
The second matrix $\Delta_{\alpha \beta}(\Omega)$ is also
hermitian. From \eqref{me97b} it follows that
\begin{equation} \hspace*{-6mm}
   \Delta^{\ast}_{\alpha \beta}(\Omega)
   = - \: \frac{1}{2i}
     \bigl[ W^{\ast}_{\alpha \beta}(\Omega) -
            W_{\beta\alpha}(\Omega)
     \bigr]
    = \frac{1}{2i}
     \bigl[ W_{\beta\alpha}(\Omega)
          - W^{\ast}_{\alpha \beta}(\Omega)
     \bigr]
   = \Delta_{\beta\alpha}(\Omega).
\label{me100} \end{equation}

Let us focus on the method of  computation of elements
$\Gamma_{\alpha \beta}(\Omega)$. As it will be shown, elements
$\Delta_{\alpha \beta}(\Omega)$ are less important.
To find $\Gamma_{\alpha\beta}$
we need quantities $W^{\ast}_{\beta \alpha}$. Conjugating
definition \eqref{me73b} we find that
\begin{align}
  W^{\ast}_{\beta\alpha}(\Omega)
  &= \left(
      \int_{0}^{\infty} d\tau ~e^{i \Omega \tau}
      ~\TTr_{B} \bigl\{ \widetilde{X}^{\dagger}_{\beta}(\tau)
         X_{\alpha} \sib \bigr\} \right)^{\ast}
\nonumber \\
  &= \int_{0}^{\infty} d\tau ~e^{ - i \Omega \tau}
   ~\TTr_{B} \bigl\{ X_{\alpha}^{\dagger}
   \widetilde{X}_{\beta}(\tau) \sib \bigr\}
\nonumber \\
  &= \int_{0}^{\infty} d\tau ~e^{ - i \Omega \tau}
     ~\TTr_{B} \bigl\{ e^{-iH_{B}\tau/\hbar} X^{\dagger}_{\alpha}
      e^{iH_{B}\tau/\hbar} X_{\beta} \sib \bigr\},
\label{me101c} \end{align}
where we used relations \eqref{me23d}, \eqref{me32} and cyclic
property of trace. Changing the integration variable
$\tau \rightarrow -\tau$, we have
\begin{equation}
   W^{\ast}_{\beta\alpha}(\Omega)
   =  \int_{-\infty}^{0} d\tau ~e^{i \Omega \tau}
     ~\TTr_{B} \bigl\{ \widetilde{X}^{\dagger}_{\alpha}(\tau)
      X_{\beta} \sib \bigr\}.
\label{me101d} \end{equation}
The integrand is identical as in the definition \eqref{me73b},
only the integration limits are different. Combining this
with \eqref{me73b}, we get
\begin{equation}
   \Gamma_{\alpha \beta}(\Omega)
   = \int_{-\infty}^{\infty} d\tau ~e^{i \Omega \tau}
     ~\TTr_{B} \bigl\{ \widetilde{X}^{\dagger}_{\alpha}(\tau)
        X_{\beta} \sib \bigr\}
   = \int_{-\infty}^{\infty} d\tau ~e^{i \Omega \tau}
     \bar{G}_{\alpha \beta}(\tau).
\label{me101g} \end{equation}
The elements $\Gamma_{\alpha \beta}(\Omega)$ are the Fourier
transforms of the corresponding correlation function of the
reservoir.

Matrix $\Delta_{\alpha \beta}(\Omega)$ does not have such
a simple representation. From the definition \eqref{me97b}
and the second relation in \eqref{me101c}
\begin{align}
  & \Delta_{\alpha \beta}(\Omega)
  = \frac{1}{2i}
    \Bigl[
       \int_{0}^{\infty} d\tau ~e^{ i \Omega \tau}
      ~\TTr_{B} \bigl\{ \widetilde{X}_{\alpha}^{\dagger}(\tau)
       X_{\beta} \sib \bigr\}
\nonumber \\
& \hspace*{40mm}
   ~-~ \int_{0}^{\infty} d\tau ~e^{ - i \Omega \tau}
      ~\TTr_{B} \bigl\{ X_{\alpha}^{\dagger}
       \widetilde{X}_{\beta}(\tau) \sib \bigr\}
    \Bigr].
\label{me101h} \end{align}

\subsection{Standard form}

Inverting relations \eqref{me97} we express elements
$W_{\alpha\beta}$ via $\Gamma_{\alpha\beta}$ and
$\Delta_{\alpha\beta}$. After simple regrouping of the terms
in Eq.\eqref{me96b} we find
\begin{align}
& \ddt \, \rho_{A}(t) \Bigl|_{d.}
  = \: \frac{1}{\hbar^{2}}
     \sum_{\Omega} \sum_{\alpha,\beta} \;
     \Gamma_{\alpha \beta}(\Omega) \:
     \Bigl\{ \: A_{\beta}(\Omega) \: \rho_{A}(t) \:
                A^{\dagger}_{\alpha}(\Omega)
\nonumber \\
& \hspace*{8mm}
- \pol \: \Bigl[ A^{\dagger}_{\alpha}(\Omega) \,
                 A_{\beta}(\Omega), \: \rho_{A}(t) \Bigr]_{+}
  - i \Delta_{\alpha \beta}(\Omega)
        \Bigl[ A^{\dagger}_{\alpha}(\Omega)
               A_{\beta}(\Omega), \: \rho_{A}(t) \Bigr]
     \: \Bigr\}.
\label{me102d} \end{align}
Let us note that the last term is a commutator, so we define
\begin{equation}
   H_{LS}
   = \frac{1}{\hbar} \sum_{\Omega} \sum_{\alpha,\beta}
     \Delta_{\alpha \beta}(\Omega)
     A^{\dagger}_{\alpha}(\Omega) A_{\beta}(\Omega).
\label{me103} \end{equation}
Taking into account hermiticity of matrix
$\Delta_{\alpha \beta}(\omega)$ (changing the names of the
summation indices when necessary)
we can easily show that the operator $H_{LS}$
is also hermitian. Returning to full master equation, that is to
Eq.\eqref{me91}, we conclude that the term containing $H_{LS}$
in \eqref{me102d} can be connected with the free hamiltonian one.
In this manner, we finally have
\begin{align}
  \ddt \, \rho_{A}(t)
  &= -\:\frac{i}{\hbar}
       \bigl[ H_{A} + H_{LS}, \: \rho_{A}(t) \bigr]
\nonumber \\
  &+~\frac{1}{\hbar^{2}}
     \sum_{\Omega} \sum_{\alpha,\beta} \;
     \Gamma_{\alpha \beta}(\Omega) \:
     \Bigl\{ \: A_{\beta}(\Omega) \: \rho_{A}(t) \:
                A^{\dagger}_{\alpha}(\Omega)
\nonumber \\
& \hspace*{40mm}
     - \pol \: \bigl[ A^{\dagger}_{\alpha}(\Omega) \:
                A_{\beta}(\Omega), \: \rho_{A}(t) \bigr]_{+}
     \Bigr\},
\label{me106} \end{align}
which coincides exactly with the standard form of the evolution
equation for the reduced density operator $\rho_{A}(t)$ which
describes the state of the subsystem $\cal{A}$ interacting with
reservoir $\cal{B}$. This allows us to be sure that hermiticity,
normalization  and positivity of the operator $\rho_{A}(t)$
are indeed ensured. Finally, let us remark that operator $H_{LS}$
which gives a contribution to the hamiltonian (unitary)
evolution, usually produces small shifts of the eigenenergies
of the subsystem $\cal{A}$. That is why, in many practical
applications, this term is simply omitted. This explains our
previous remark that matrix $\Delta_{\alpha \beta}$
is less important than $\Gamma_{\alpha \beta}$.
Obviously one can  construct operator $H_{LS}$ and investigate
its influence on the unperturbed energy levels of the subsystem
$\cal{A}$. Small energy shifts of eigenenergies of subsystem
$\cal{A}$ are qualitatively similar to the well-known
Lamb shifts, which clarifies the employed notation.

The obtained master equation \eqref{me106} is an operator one.
In practice, we frequently need an equation of motion for the
matrix elements of the reduced density operator $\rho_{A}(t)$.
It seems to be natural to use the energy representation,
that is to consider matrix elements of $\rho_{A}(t)$
calculated in the basis $\left\{ \ket{a} \right\}$ of the
eigenstates of the free hamiltonian $H_{A}$ (see Eq.\eqref{me44}).
This will be done in the next section.

\subsection{Energy representation}

When analyzing master equation in the basis of the eigenstates
of free hamiltonian we need to be careful. The reason is that
the commutator in \eqref{me106} contains an additional
term, namely the Lamb-shift hamiltonian. One may argue that
this changes the hamiltonian and a new basis should be found
(a basis in which $H_{A}+H_{LS}$ is diagonal). We will,
however, proceed in the spirit of the perturbative approach.
We will treat $H_{LS}$ as a small perturbation which, at most,
will yield small energy shifts. Therefore, the set
$\{\ket{a}\}$ of eigenstates of the unperturbed hamiltonian
$H_{A}$ can be used as complete and orthonormal basis.
Working within this scheme, we can easily construct master
equation (equation of motion) for matrix elements of the
density operator for subsystem $\cal{A}$. We will
suppress the index $A$ since it should lead to no
misunderstanding. Taking matrix elements $\rho_{ab}(t) = %
\elm{a}{\rho_{A}(t)}{b}$ and expanding the anticommutator
term we obtain
\begin{align}
   \ddt \: \rho_{ab}(t)
   =& - \: \frac{i}{\hbar}
   \bra{a} \bigl[ H_{A} + H_{LS}, \; \rho(t) \bigr] \ket{b}
\nonumber \\
& +~ \frac{1}{\hbar^{2}} \sum_{\Omega} \sum_{\alpha,\beta} \:
     \Gamma_{\alpha \beta}(\Omega) \:
     \Bigl\{ \: \bra{a} A_{\beta}(\Omega) \: \rho(t) \:
           A^{\dagger}_{\alpha}(\Omega) \ket{b}
\nonumber \\
& \hspace*{-5mm}
  - \pol \:
    \bra{a} A^{\dagger}_{\alpha}(\Omega) \: A_{\beta}(\Omega) \:
    \rho(t) \ket{b}
  - \pol \:
    \bra{a} \rho(t) \: A^{\dagger}_{\alpha}(\Omega) \:
    A_{\beta}(\Omega) \ket{b}
    \Bigr\}.
\label{me112} \end{align}
The last three terms constitute a so-called dissipative term
(or dissipator) and we will concentrate on its form.
First, we use expressions \eqref{me45}, \eqref{me47}
for operators $ A_{\alpha}(\Omega)$ and
$A^{\dagger}_{\alpha}(\Omega)$. Then we consider three matrix
elements. Necessary computations in the basis of eigenstates
of free hamiltonian $H_{A}$ are simple though a bit tedious,
in some cases a suitable changes of summation indices is
necessary. The results of these calculations are as follows
\begin{subequations} \label{me117}
\begin{align}
& \bra{a}A_{\beta}(\Omega)\:\rho(t) \:
          A^{\dagger}_{\alpha}(\Omega)\ket{b} =
\nonumber \\[2mm]
& \hspace*{9mm}
   =~\sum_{m,n}
     \delta(\omega_{ma} - \Omega) \: \delta(\omega_{nb} - \Omega)
     \elm{a}{A_{\beta}}{m} \elm{n}{A^{\dagger}_{\alpha}}{b} \:
     \rho_{mn}(t),
\label{me117a} \end{align}
\begin{align}
& \bra{a}A^{\dagger}_{\alpha}(\Omega)\:
          A_{\beta}(\Omega)\:\rho(t)\ket{b} =
\nonumber \\[2mm]
& \hspace*{9mm}
   = \sum_{m,n}
     \delta(\omega_{an} - \Omega) \: \delta(\omega_{mn} - \Omega)
     \elm{a}{A^{\dagger}_{\alpha}}{n} \elm{n}{A_{\beta}}{m} \:
     \rho_{mb}(t),
\label{me117b} \end{align}
\begin{align}
& \bra{a}\rho(t)\:A^{\dagger}_{\alpha}(\Omega)\:
                   A_{\beta}(\Omega)\ket{b} =
\nonumber \\[2mm]
& \hspace*{9mm}
   = \sum_{m,n}
     \delta(\omega_{mn} - \Omega) \: \delta(\omega_{bn} - \Omega)
     \elm{m}{A^{\dagger}_{\alpha}}{n} \elm{n}{A_{\beta}}{b} \:
     \rho_{am}(t).
\label{me117c}
\end{align} \end{subequations}
The computed matrix elements are plugged into equation \eqref{me112}
and summation over frequency $\Omega$ is performed.
After some regrouping we find that
\begin{align}
& \hspace*{-8mm}
   \ddt \: \rho_{ab}(t) \Bigr|_{d.}
   = \frac{1}{\hbar^{2}}
     \sum_{\alpha,\beta} \sum_{m,n}
     \Bigr\{ \Gamma_{\alpha \beta}(\omega_{ma}) \:
             \delta(\omega_{nb} - \omega_{ma})
             \elm{a}{A_{\beta}}{m}
             \elm{b}{A_{\alpha}}{n}^{\ast} \: \rho_{mn}(t)
\nonumber \\
& \hspace*{13mm}
   -~\pol \: \Gamma_{\alpha \beta}(\omega_{an}) \:
     \delta(\omega_{mn} - \omega_{an})
     \elm{n}{A_{\beta}}{m}
     \elm{n}{A_{\alpha}}{a}^{\ast} \: \rho_{mb}(t)
\nonumber \\[2mm]
& \hspace*{13mm}
   -~\pol \: \Gamma_{\alpha \beta}(\omega_{mn}) \:
     \delta(\omega_{bn} - \omega_{mn})
     \elm{n}{A_{\beta}}{b}
     \elm{n}{A_{\alpha}}{m}^{\ast} \: \rho_{am}(t)
     \Bigr\}.
\label{me118b} \end{align}
Going further, we use the evenness of Kronecker delta in the first
term, while the presence of the deltas in the second and third terms
allows us to change arguments in the elements of matrix
$\Gamma_{\alpha \beta}$. Next, we denote
\begin{equation}
   K(am,bn)
   ~=~ \frac{1}{\hbar^{2}}
       \sum_{\alpha,\beta} \Gamma_{\alpha \beta}(\omega_{ma})
       \elm{a}{A_{\beta}}{m}
       \elm{b}{A_{\alpha}}{n}^{\ast},
\label{me119} \end{equation}
Due to these, we rewrite formula \eqref{me118b} as
\begin{align}
& \ddt \: \rho_{ab}(t) \Bigr|_{d.}
   ~=~ \sum_{m,n} \: \delta(\omega_{ma} - \omega_{nb}) \;
       K(am,bn) \; \rho_{mn}(t)
\nonumber \\
& \hspace*{30mm}
   -~\pol \:
     \sum_{m,n} \: \delta(\omega_{mn} - \omega_{an}) \;
     K(nm,na) \; \rho_{mb}(t)
\nonumber \\[2mm]
& \hspace*{40mm}
   -~\pol \:
     \sum_{m,n} \: \delta(\omega_{bn} - \omega_{mn}) \;
     K(nb,nm) \; \rho_{am}(t).
\label{me120} \end{align}
Let us note the specific symmetry of this expression.
Further analysis depends on whether the eigenfrequencies
of the hamiltonian $H_{A}$ are degenerate or not.
We also note that Kronecker deltas in the second and third
terms are correspondingly given as
$\delta(\omega_{mn} - \omega_{an}) = \delta(\omega_{ma})$
and
$\delta(\omega_{bn} - \omega_{mn}) = \delta(\omega_{bm})$,
which allows one to perform summation over $n$.
However, one has to be careful because eigenfrequencies
$\omega_{n}$ can be degenerate.

\subsection{Degenerate eigenfrequencies}

To account for the possible degeneracies, let us write the
hamiltonian of the considered system $\cal{A}$
in the following form
\begin{equation}
   H_{A}
   = \sum_{N} \hbar \omega_{N}
     \sum_{n=1}^{g_{N}} \ket{Nn} \bra{Nn},
\label{me130} \end{equation}
where $N$ is the main quantum number which distinguishes
energy levels (energy multiplets), while $n=1,2,\ldots,g_{N}$,
are subsidiary quantum numbers. Is is obvious that
$\omega_{N} \neq \omega_{M}$ for $N \neq M$.
Certainly, the nondegenerate case follows immediately and
it corresponds to $g_{N} \equiv 1$, then subsidiary quantum
numbers are unnecessary and can be simply suppressed.

In the degenerate case single indices appearing in equation
\eqref{me120} must be replaced by corresponding pairs,
for example $a \rightarrow Aa$. Equation \eqref{me120}
is now rewritten as
\begin{align}
& \ddt \: \rho_{AaBb}(t) \Bigr|_{d.}
   ~=~ \sum_{Mm} \sum_{Nn} \: \delta(\omega_{MA} - \omega_{NB}) \;
       K(AaMm,BbNn) \; \rho_{MmNn}(t)
\nonumber \\
& \hspace*{10mm}
   -~\pol \:
     \sum_{Mm} \sum_{Nn} \: \delta(\omega_{MN} - \omega_{AN}) \;
     K(NnMm,NnAa) \; \rho_{MmBb}(t)
\nonumber \\[2mm]
& \hspace*{10mm}
   -~\pol \:
     \sum_{Mm} \sum_{Nn} \: \delta(\omega_{BN} - \omega_{MN}) \;
     K(NnBb,NnMm) \; \rho_{AaMm}(t).
\label{me131} \end{align}
As already noted, one immediately sees that
$\delta(\omega_{MN} - \omega_{AN}) =%
\delta(\omega_{MA}) = \delta_{MA}$ and similarly
$\delta(\omega_{BN} - \omega_{MN}) = \delta_{MB}$,
where the last deltas are the simple Kronecker ones.
The sum over $M$ in the second term is trivial. We put $M=A$
and we "land within multiplet $A$", hence we change $m=a''$.
Analogously, in the second term  $M=B$ and $m=b''$.
Therefore, we have
\begin{align}
& \ddt \: \rho_{AaBb}(t) \Bigr|_{d.}
   ~=~ \sum_{Mm} \sum_{Nn} \: \delta(\omega_{MA} - \omega_{NB}) \;
       K(AaMm,BbNn) \; \rho_{MmNn}(t)
\nonumber \\
& \hspace*{40mm}
   -~\pol \:
     \sum_{Nn} \sum_{a''} K(NnAa'',NnAa) \; \rho_{Aa''Bb}(t)
\nonumber \\[2mm]
& \hspace*{40mm}
   -~\pol \:
     \sum_{Nn} \sum_{b''} K(NnBb,NnBb'') \; \rho_{AaBb''}(t).
\label{me132} \end{align}
In two last terms matrix elements do not depend on
quantum numbers $Nn$, hence we can denote
\begin{equation}
   \kappa(Aa,Bb) = \sum_{Nn} K(NnAa,NnBb).
\label{me133} \end{equation}
This allows us to write equation \eqref{me133} in the form
\begin{align}
& \ddt \: \rho_{AaBb}(t) \Bigr|_{d.}
   ~=~ \sum_{Mm} \sum_{Nn} \: \delta(\omega_{MA} - \omega_{NB}) \;
       K(AaMm,BbNn) \; \rho_{MmNn}(t)
\nonumber \\
& \hspace*{10mm}
   -~\pol \:\sum_{a''}\kappa(Aa'',Aa)\;\rho_{Aa''Bb}(t)
  ~-~\pol \:\sum_{b''}\kappa(Bb,Bb'')\;\rho_{AaBb''}(t).
\label{me135} \end{align}

Let us consider this equation in some more detail.
First, we take $A=B$ (and correspondingly $b \rightarrow a'$).
This yields the equation of motion for "quasi-population"
-- matrix elements taken within just one energy multiplet.
Then, the first term in right-hand side contains
$\delta(\omega_{MA} - \omega_{NA}) = \delta(\omega_{MN})%
=\delta_{MN}$. The sum over $N=M$ is trivial ($n \rightarrow m'$)
and we have
\begin{align}
& \ddt \: \rho_{AaAa'}(t) \Bigr|_{d.}
   ~=~ \sum_{Mmm'} K(AaMm,Aa'Mm') \; \rho_{MmMm'}(t)
\nonumber \\
& \hspace*{10mm}
   -~\pol \:\sum_{a''}\kappa(Aa'',Aa)\;\rho_{Aa''Aa'}(t)
  ~-~\pol \:\sum_{a''}\kappa(Aa',Aa'')\;\rho_{AaAa''}(t).
\label{me141} \end{align}
This equation connects "quasi-populations" with other ones.
The first sum contains the term with $A=M$ and this term represents
elastic (energy conserving) processes. The remaining terms
(with $M \neq A$) corresponding to nonelastic transitions.
In this case, the environment $\cal{B}$ serves as a reservoir
which gives or absorbs the energy. The terms in the second line
describe the "escape" from multiplet $A$ to other ones.

To discuss coherences, we assume $A \neq B$, which implies
$\omega_{A} \neq \omega_{B}$. The Kronecker delta in \eqref{me135}
can be rewritten as
$\delta(\omega_{MA} - \omega_{NA}) = \delta(\omega_{MN} - \omega_{AB})$.
Since $\omega_{AB} \neq 0$, we also get $\omega_{MN} \neq 0$.
If we assume that all energy distances are different
(that is $\omega_{AB} \neq \omega_{MN}$ for different
pairs $A,B \neq M,N$) the considered delta can give unity
only when $A=M$ and $B=N$ (which entails $m \rightarrow a'$
and $n \rightarrow b').$ Then, Eq. \eqref{me135} reduces to
\begin{align}
& \ddt \: \rho_{AaBb}(t) \Bigr|_{d.}
   ~=~ \sum_{a'} \sum_{b'} \: K(AaAa',BbBb') \; \rho_{Aa'Bb'}(t)
\nonumber \\
& \hspace*{11mm}
   -~\pol \:\sum_{a''}\kappa(Aa'',Aa)\;\rho_{Aa''Bb}(t)
  ~-~\pol \:\sum_{b''}\kappa(Bb,Bb'')\;\rho_{AaBb''}(t).
\label{me144} \end{align}
So the coherences between two multiplets $A \neq B$ couple
only with coherences from just these multiplets.

Obviously for the nondegenerate case  "small" indices play no role
-- they can be suppressed. Then, instead of equation \eqref{me141}
for "quasi-populations" we get an equation for genuine
populations
\begin{equation}
   \ddt \: \rho_{AA}(t) \Bigr|_{d.}
   ~=~ \sum_{M} K(AM,AM) \; \rho_{MM}(t)
   -~\kappa(A,A) \; \rho_{AA}(t).
\label{me147} \end{equation}
Similarly for coherences, Eq. \eqref{me144} yields
\begin{equation}
   \ddt \: \rho_{AB}(t) \Bigr|_{d.}
   ~=~ K(AA,BB) \; \rho_{AB}(t)
   -~ \frac{1}{2} \:\biggl[ \kappa(A,A) + \kappa(B,B) \biggr]
   \;\rho_{AB}(t).
\label{me148} \end{equation}
These examples indicate that ME for matrix elements
of the reduced density operator possess quite a specific symmetry
which probably can be further investigated. This, however,
goes beyond the scope of the present work.

\section{Auxiliary sections}

\subsection{Preservation of normalization}

Any density operator, so also the reduced one for subsystem
${\cal A}$ must be normalized, that is, we require that
$\TTr_{A} \{\rho_{A}(t)\}=1$.
This has a simple consequence
\begin{equation}
   \ddt \TTr_{A} \{ \rho_{A}(t) \}
   = \TTr_{A} \left\{ \frac{d\,\rho_{A}(t)}{dt} \right\}
   = 0.
\label{mea02} \end{equation}
Clearly the hamiltonian part (the commutator) preserves the trace,
which follows from cyclic property. Hence we must check the
second -- dissipative part of our ME. One may ask at which
stage of our derivation such a check should be made.
In principle, this can be done at any stage. In this section
we shall do so twice. Once for standard form \eqref{me106},
and for ME \eqref{me135} in the energy basis.

\subsubsection*{Standard form}

Taking ME in its standard form \eqref{me106} we need to compute
the following trace
\begin{align}
& t_{S}
  ~= \TTr_{A}
     \Bigl\{ \sum_{\Omega} \sum_{\alpha,\beta}
     \Bigl[ \: \Gamma_{\alpha \beta}(\Omega) \:
                A_{\beta}(\Omega) \: \rho_{A}(t) \:
                A^{\dagger}_{\alpha}(\Omega)
\nonumber \\
& \hspace*{27mm}
 ~-~ \pol \: \Gamma_{\alpha \beta}(\Omega) \:
      A^{\dagger}_{\alpha}(\Omega) \:
      A_{\beta}(\Omega) \: \rho_{A}(t)
\nonumber \\
& \hspace*{42mm}
 ~-~ \pol \: \Gamma_{\alpha \beta}(\Omega) \:
      \rho_{A}(t) \: A^{\dagger}_{\alpha}(\Omega) \:
      A_{\beta}(\Omega)
      \Bigr] \Bigr\},
\label{mea03} \end{align}
and show that it vanishes, ie., $t_{S}=0$. The trace is a linear
operation, so then
\begin{align}
& t_{S}
  ~= \sum_{\Omega} \sum_{\alpha,\beta}
     \Bigl[ \: \Gamma_{\alpha \beta}(\Omega) \:
     \TTr_{A} \Bigl\{A_{\beta}(\Omega) \: \rho_{A}(t) \:
           A^{\dagger}_{\alpha}(\Omega) \Bigr\}
\nonumber \\
& \hspace*{27mm}
 ~-~ \pol \: \Gamma_{\alpha \beta}(\Omega) \:
     \TTr_{A} \Bigl\{ A^{\dagger}_{\alpha}(\Omega) \:
           A_{\beta}(\Omega) \: \rho_{A}(t) \Bigr\}
\nonumber \\[2mm]
& \hspace*{42mm}
 ~-~ \pol \: \Gamma_{\alpha \beta}(\Omega) \:
     \TTr_{A} \Bigl\{ \rho_{A}(t) \:
        A^{\dagger}_{\alpha}(\Omega) \: A_{\beta}(\Omega)
     \Bigr\} \Bigr].
\label{mea04} \end{align}
Cyclic property allows one to see that all three traces are equal.
Therefore, $t_{S}=0$ and we conclude that preservation of
the normalization for ME in the standard form is proved.

\subsubsection*{ME in energy basis}

In this case we check the trace preservation for
Eq.\eqref{me141}, with $a'=a$. We need to compute
\begin{align}
& \hspace*{-8mm}
   t_{S}
   ~=~ \sum_{Aa} \sum_{Mm,m'} \:
       K(AaMm,AaMm') \; \rho_{MmMm'}(t)
\nonumber \\
& \hspace*{-2mm}
   -~\pol \:
     \sum_{Aa} \sum_{a''}\kappa(Aa'',Aa)\;\rho_{Aa''Aa}(t)
  ~-~\pol \:
     \sum_{Aa} \sum_{a''}\kappa(Aa,Aa'')\;\rho_{AaAa''}(t).
\label{mea06} \end{align}
In the first term we use definition of the parameter
$\kappa$ (see \eqref{me133}). In the two next ones
we notice that indices $a$ and $a''$ concern the same
multiplet $A$, so the summation range is also the same.
We can interchange $a \rightarrow a''$ and obtain
\begin{align}
& \hspace*{-8mm}
   t_{S}
   ~=~ \sum_{Mm,m'} \: \kappa(Mm,Mm') \; \rho_{MmMm'}(t)
\nonumber \\
& \hspace*{-2mm}
   -~\pol \:
     \sum_{Aa} \sum_{a''}\kappa(Aa,Aa'')\;\rho_{AaAa''}(t)
  ~-~\pol \:
     \sum_{Aa} \sum_{a''}\kappa(Aa,Aa'')\;\rho_{AaAa''}(t).
\label{mea07} \end{align}
The second and third terms are identical and cancel out with
the first one (names of summation indices are irrelevant).
We have shown that in the energetic basis the trace of the
reduced density operator for subsystem $\cal{A}$ is preserved.
In other words, the derived ME preserves normalization.

\subsection{Hermiticity of the reduced density operator}

The next necessary property of any density operator is its
hermiticity. If the equation of motion for $\rho_{A}^{\dagger}(t)$
is identical with the similar equation for $\rho_{A}(t)$, then
the same equations must yield the same solutions, this means
that $\rho_{A}^{\dagger}(t)=\rho_{A}(t)$.
Free evolution is given by the hamiltonian term
$(-i/\hbar)\left[ H_{A}+H_{LS}, \: \rho_{A}(t) ~\right]$
which poses no problems due to the commutator properties.
One needs to investigate the dissipative part of ME.
As in the previous section we perform such a check for
ME in standard form and for the one in energy basis.

\subsubsection*{Standard form}

We take the hermitian conjugate of the dissipative part
of ME
\begin{align}
& \ddt \, \rho_{A}^{\dagger}(t) \Bigr|_{d.}
  ~= +~\frac{1}{\hbar^{2}}
       \sum_{\Omega} \sum_{\alpha,\beta}
       \Bigl\{ \: \Gamma_{\alpha \beta}^{\ast}(\Omega) \:
          A_{\alpha}(\Omega) \: \rho_{A}^{\dagger}(t) \:
          A_{\beta}(\Omega)
\nonumber \\
& \hspace*{48mm}
     ~-~ \pol \: \Gamma_{\alpha \beta}^{\ast}(\Omega) \:
         \bigl[ A^{\dagger}_{\beta}(\Omega) \:
                A_{\alpha}(\Omega), \: \rho_{A}^{\dagger}(t)
         \bigr]_{+}
         \Bigr\},
\label{meb01} \end{align}
because conjugate anticommutator is equal to the anticommutator
of conjugated operators. We know (see \eqref{me98}) that
matrix $\Gamma_{\alpha \beta}$ is hermitian. Interchanging the
indices $\alpha \leftrightarrow \beta$ we get
\begin{align}
& \ddt \, \rho_{A}^{\dagger}(t) ) \Bigr|_{d.}
  ~= +~\frac{1}{\hbar^{2}}
       \sum_{\Omega} \sum_{\alpha,\beta}
       \Bigl\{ \: \Gamma_{\alpha \beta}(\Omega) \:
          A_{\beta}(\Omega) \: \rho_{A}^{\dagger}(t) \:
          A_{\alpha}(\Omega)
\nonumber \\
& \hspace*{48mm}
     ~-~ \pol \: \Gamma_{\alpha \beta}(\Omega) \:
         \bigl[ A^{\dagger}_{\alpha}(\Omega) \:
                A_{\beta}(\Omega), \: \rho_{A}^{\dagger}(t)
         \bigr]_{+}
         \Bigr\},
\label{meb02} \end{align}
We see that the equation of motion for $\rho_{A}^{\dagger}$
is formally identical with standard form \eqref{me106} of ME.
Thus, $\rho_{A}^{\dagger} = \rho_{A}$ -- hermiticity
is preserved.

\subsubsection*{ME in energetic basis}

Hermiticity of the density operator means that
$\rho_{AaBb} = \rho_{BbAa}^{\ast}$. It entails, that the equation
of motion for the element $\rho_{BbAa}^{\ast}$ must be the same
as for $\rho_{AaBb}$. Starting from ME \eqref{me135} we look for
a corresponding equation for $\rho_{BbAa}^{\ast}$.
First we need to change the indices in \eqref{me135} (remembering
that corresponding changes must be made for summation indices
in all terms) and then we perform complex conjugation. In this
manner we find
\begin{align}
& \ddt \: \rho_{BbAa}^{\ast}(t) \Bigr|_{d.}
   ~=~ \sum_{Mm} \sum_{Nn} \: \delta(\omega_{MB} - \omega_{NA}) \;
       K^{\ast}(BbMm,AaNn) \; \rho_{MmNn}^{\ast}(t)
\nonumber \\
& \hspace*{8mm}
   -~\pol \:\sum_{b''}\kappa^{\ast}(Bb'',Bb)\;
     \rho_{Bb''Aa}^{\ast}(t)
  ~-~\pol \:\sum_{a''}\kappa^{\ast}(Aa,Aa'')\;
     \rho_{BbAa''}^{\ast}(t).
\label{meb03c} \end{align}
Next, we need to consider the conjugated quantities $K^{\ast}$
and $\kappa^{\ast}$. By definition \eqref{me119}
\begin{align}
& \delta(\omega_{MA}-\omega_{NB})
   K(AaMm,BbNn) ~=
\nonumber \\
& \hspace*{18mm}
   =~\frac{1}{\hbar^{2}}
     \sum_{\alpha,\beta} \Gamma_{\alpha \beta}(\omega_{MA})
     \elm{Aa}{A_{\beta}}{Mm}
     \elm{Bb}{A_{\alpha}}{Nn}^{\ast}.
\label{meb04} \end{align}
We take complex conjugates, use hermiticity of matrix
$\Gamma_{\alpha\beta}$ and we notice that the presence of the
Kronecker delta allows the change of the argument in $\Gamma$.
Interchanging the summation indices $\alpha \leftrightarrow \beta$
we have
\begin{align}
& \delta(\omega_{MA}-\omega_{NB})
   K^{\ast}(AaMm,BbNn) ~=
\nonumber \\
& \hspace*{18mm}
   =~\frac{1}{\hbar^{2}}
     \sum_{\alpha,\beta} \Gamma_{\alpha \beta}(\omega_{NB})
     \elm{Bb}{A_{\beta}}{Nn}
     \elm{Aa}{A_{\alpha}}{Mm}^{\ast}.
\label{meb05} \end{align}
Comparing this relation with definition \eqref{me119}
we see that
\begin{equation}
   \delta(\omega_{MA}-\omega_{NB}) K^{\ast}(AaMm,BbNn)
   ~=~\delta(\omega_{MA}-\omega_{NB}) K(BbNn,AaMm).
\label{meb06} \end{equation}
Next, we deal with the parameter $\kappa^{\ast}(Aa'',Aa)$.
In the above relation we substitute $Aa \rightarrow Nn$,
$Mm \rightarrow Aa''$, $Bb \rightarrow Nn$ and
$Nn \rightarrow Aa$. Then
\begin{equation}
   \delta(\omega_{AN}-\omega_{AN}) K^{\ast}(NnAa'',NnAa)
   ~=~\delta(\omega_{AN}-\omega_{AN}) K(NnAa,NnAa'').
\label{meb08} \end{equation}
Obviously Kronecker deltas are equal to one, so they are
unimportant. Using this result in the definition \eqref{me133}
of the parametr $\kappa$ we get
\begin{align}
   \kappa^{\ast}(Aa'',Aa)
   &~=~ \sum_{Nn} K^{\ast}(NnAa'',NnAa)
\nonumber \\
   &~=~ \sum_{Nn} K(NnAa,NnAa'')
   ~=~ \kappa(Aa,Aa'').
\label{meb10} \end{align}
Returning to the analysis of formula \eqref{meb03c}, we use
the proven relations \eqref{meb06} and \eqref{meb10}.
At the same time, in the first term in the right hand side
we interchange the summation indices $Mm \leftrightarrow Nn$.
Moreover, we recall that Kronecker delta is even. Thus, we have
\begin{align}
& \ddt \: \rho_{BbAa}^{\ast}(t) \Bigr|_{d.}
   ~=~ \sum_{Mm} \sum_{Nn} \:
       \delta(\omega_{MA} - \omega_{NB}) \;
       K(AaMm,BbNn) \; \rho_{NnMm}^{\ast}(t)
\nonumber \\
& \hspace*{8mm}
   -~\pol \:\sum_{b''}\kappa(Bb,Bb'')\;
     \rho_{Bb''Aa}^{\ast}(t)
  ~-~\pol \:\sum_{a''}\kappa(Aa'',Aa)\;
     \rho_{BbAa''}^{\ast}(t).
\label{meb12} \end{align}
Comparing this result with Eq.\eqref{me135} we find that
when we replace $\rho_{AaBb}$ by $\rho_{BbAa}^{\ast}$
(consequently in all the terms) then we will arrive
at \eqref{meb12}. This equation is formally identical
with Eq.\eqref{me135}, hence $\rho = \rho^{\dagger}$,
what we intended to show. Our ME preserves hermiticity
of the reduced density operator of subsystem $\cal{A}$.

\subsection{Reduced density operator in interaction picture}

In this section we will show that the reduction of the density
operator is invariant with respect to the choice of the picture.
The definition \eqref{ii07} of the reduced density operator
and \eqref{me06} for the transformation to the interaction picture
imply that the operator $\rho_{A}(t)$ (in the Schr{\"{o}}dinger
picture) is expressed as
\begin{equation}
   \rho_{A}(t)
   ~=~ \TTr_{B} \{ \rho_{AB}(t) \}
   ~=~ \TTr_{B}
       \left\{ \:  e^{-iH_{0}t/\hbar} \; \inro_{AB}(t) \;
                   e^{ iH_{0}t/\hbar} \: \right\}.
\label{mec09} \end{equation}
We note that the free evolution of each of the subsystems
written as
\begin{equation}
    \ket{\varphi_{A}(0)} \otimes \ket{\psi_{B}(0)}
    \longrightarrow
    \ket{\varphi_{A}(t)} \otimes \ket{\psi_{B}(t)},
\label{mec10a} \end{equation}
can be expressed with the aid of the operator
\begin{equation}
   \exp \left( - \frac{i}{\hbar} H_{0}t \right)
   ~=~ \exp \left( - \frac{i}{\hbar} H_{A}t \right) \otimes
       \exp \left( - \frac{i}{\hbar} H_{B}t \right),
\label{mec10b} \end{equation}
because both hamiltonians are fully independent and commute.
In Eq.\eqref{mec09} we compute the trace only over reservoir
variables, so we can write
\begin{equation}
   \rho_{A}(t)
    = e^{-iH_{A}t/\hbar} \;
      \TTr_{B} \left\{ \:e^{-iH_{B}t/\hbar} \;
           \inro_{AB}(t) \; e^{ iH_{B}t/\hbar} \: \right\}
       e^{ iH_{A}t/\hbar}.
\label{mec11} \end{equation}
Cyclic property of the trace yields
\begin{equation}
   e^{ iH_{A}t/\hbar} \; \rho_{A}(t) \; e^{-iH_{A}t/\hbar}
    ~=~ \TTr_{B} \left\{ \: \inro_{AB}(t) \: \right\}.
\label{mec12} \end{equation}
Left hand side represents the reduced density operator
in the interaction picture (it depends solely on the variables
of the subsystem ${\cal A}$). Hence, we have
\begin{equation}
  \inro_{A}(t)
  ~=~ \TTr_{B} \left\{ \: \inro_{AB}(t) \: \right\}.
\label{mec13} \end{equation}
This is formally identical with the definition of the
reduced density operator in the Schr{\"{o}}dinger picture.
The relation between the reduced density and the total one
is the same in both pictures. In other words, reduction of the
operatora $\rho_{A}(t)= \TTr_{B}\left\{\:\rho_{AB}(t)\:\right\}$
is invariant with respect to the change of the  pictures.

\subsection{Existence of two time scales. %
Discussion of approximations}

In this section we follow and (hopefully) try to to elucidate
the discussion presented by Cohen-Tannoudji {\em et al} \cite{coh}.
The previously given remarks apply also here, so we again
refrain from frequent citations.

\subsubsection*{Order of magnitude of time $T_{A}$}

The key role in our considerations is played by the assumption
\eqref{me18}. This is the requirement
\begin{equation}
   \tau_{B} \ll \Delta t \ll T_{A}.
\label{med01} \end{equation}
In other words we assume that there exist two, quite distinct,
time scales. Firstly, let us try to estimate the time $T_{A}$
which characterizes the evolution of system ${\cal A}$
due to the interaction with reservoir. To find such an estimate
we use Eq. \eqref{me41a}, that is
\begin{align}
& \Delta \, \inro_{A}(t)
    ~=~ \left( \: \frac{1}{i\hbar}\: \right)^{2}
        \int_{t}^{t+\Delta t} dt_{1}
        \int_{t}^{t_{1}} dt_{2}
       ~\TTr_{B} \biggl\{
\nonumber \\
& \hspace*{45mm}
       \left[ \: \widetilde{V}_{AB}(t_{1}), \;
       \bigl[ \: \widetilde{V}_{AB}(t_{2}),
      ~\inro_{A}(t) \otimes \sib\:
       \bigr] \right] \biggl\},
\label{med02} \end{align}
where we employed the introduced properties of the reservoir.
We also recall that the main contribution to the integrals comes
from a thin belt (of width $\tau_{B}$ lying below the diagonal
$t_{1} = t_{2}$, see Fig. \ref{xmerys01} and its discussion).
This allows us to estimate the integrand as follows
\begin{equation}
   ~\TTr_{B} \left[ \: \widetilde{V}_{AB}(t_{1}), \;
    \bigl[ \: \widetilde{V}_{AB}(t_{2}),
   ~\inro_{A}(t) \otimes \sib\:
    \bigr] \right]
    ~\sim~ \inro_{A} \:
     \TTr_{B} \left\{ \: \widetilde{V}^{2} \sib\: \right\}
    ~=~ V^{2} \inro_{A}.
\label{med03} \end{equation}
Hence, left hand side of Eq.\eqref{med02} is estimated by
\begin{equation}
  \frac{\Delta \inro_{A}}{\Delta t}
  ~\sim~ \frac{1}{\hbar^{2}} \; \tau_{B} \: V^{2} \inro_{A},
\label{med04} \end{equation}
because the area of the integration region is estimated
by the product $\tau_{B} \Delta t$ (area of the belt under
the diagonal $t_{1}=t_{2}$). Introduced parameter $V$
characterizes the "strength" of the interaction between
the reservoir and system ${\cal A}$. The factor which multiplies
$\inro_{A}$ in \eqref{med04} has (according to \eqref{me17})
the sense of the inverse of time $T_{A}$.
Therefore, we obtain an estimate
\begin{equation}
  \frac{1}{T_{A}}
  ~\sim~ \frac{V^{2} \tau_{B}}{\hbar^{2}},
  \hspace*{12mm} \mathrm{or} \hspace*{12mm}
  T_{A} ~\sim~ \frac{\hbar^{2}}{V^{2} \tau_{B}}.
\label{med05} \end{equation}

\subsubsection*{Condition for existence of two time scales}

What is the condition of the existence of two time scales?
The estimate of $T_{A}$ given in \eqref{med05} allows us to
find such a condition. Let us look upon condition
$\tau_{B} \ll T_{A}$ more carefully and introduce the estimate
\eqref{med05}. This yields
\begin{equation}
   \tau_{B} \ll \frac{\hbar^{2}}{V^{2} \tau_{B}}
   \hspace*{12mm} \Longrightarrow \hspace*{12mm}
   \frac{V \tau_{B}}{\hbar} \ll 1.
\label{med06} \end{equation}
The last inequality is the sought condition of existence of two
time scales. If we denote $\Omega_{AB} = V/\hbar$, then we can
write  $\Omega_{AB} \tau_{B} \ll 1$. So the interaction
must be characterized by such Bohr frequency $\Omega_{AB}$
that during the time interval of magnitude of $\tau_{B}$
its influence on system ${\cal A}$ is negligibly small.

\subsubsection*{Justification of weak coupling approximation}

We already mentioned (see the discussion of Eq.\eqref{me16}),
that it is possible to iterate von Neumann equation -- accounting
for higher order corrections would increase accuracy.
We can estimate these higher order terms in the same manner
as done just above. For example, for the third order term we
have
\begin{equation}
  \frac{\Delta \inro_{A}}{\Delta t} \Bigl|^{(3)}
  ~\sim~ \frac{V^{3}}{\hbar^{3}} \; \tau_{B}^{2} \: \inro_{A},
\label{med07} \end{equation}
because times $t_{1}$, $t_{2}$ and $t_{3}$ must be close
to each other (with accuracy of the order of $\tau_{B}$).
Then the region of integration has volume of the order of
$\tau_{B}^{2} \Delta t$. Due to Eq.\eqref{med05} we get
\begin{equation}
  \frac{\Delta \inro_{A}}{\Delta t} \Bigl|^{(3)}
  ~\sim~ \frac{V \tau_{B}}{\hbar} \cdot \frac{1}{T_{A}} \;
  \inro_{A}
  ~\ll~ \frac{1}{T_{A}} \: \inro_{A},
\label{med08} \end{equation}
where the last inequality follows from \eqref{med06}.
The obtained estimate shows that the third order iteration
(and similarly higher ones) are indeed negligible. Obviously this
holds provided the condition \eqref{med06} is true and ensures
the existence of two distinct time scales.

\subsubsection*{Neglecting $\rho_{corel}$ (Born approximation)}

Moving from Eq.\eqref{me16} to \eqref{me41a} we have neglected
initial correlations between systems ${\cal A}$ and ${\cal B}$.
These correlations built up at earlier moments $t' < t$.
This corresponds to the assumption that at some earlier
moment $t_{0}$ ($t_{0} < t$) both systems were uncorrelated.
This happens, for example, when the interaction was switched on
at an instant $t_{0}$. So the correlations described by
$\inro_{corel}$ need time $t-t_{0}$ to appear.
If the correlations exist ($\varrho_{corel} \neq 0$)
then averaging of the term linear in interaction (as in
expression \eqref{me40}) would not give zero.
Hence, $\varrho_{corel} \neq 0$ would result in the appearance
of the linear term. Moreover, this would also automatically
modify the quadratic term in \eqref{me41c}. Let us estimate
the magnitude of this modification (which is due to earlier
interaction)
\begin{equation}
  \frac{\Delta \inro_{A}}{\Delta t}
  ~\sim~ \frac{1}{\Delta t} \left( \frac{1}{i\hbar} \right)^{2}
  \int_{-\infty}^{t} dt_{1} \int_{t}^{t+\Delta t} dt_{2}
  ~\qav{\widetilde{V}_{AB}(t_{1}) \widetilde{V}_{AB}(t_{2})}_{B}.
\label{med09} \end{equation}
The integrand contains correlation functions of the reservoir.
Hence the integrand would be practically zero for
$|t_{1}-t_{2}| \geq \tau_{B}$. The integration runs
effectively from $t-\tau_{B}$ to $t+\tau_{B}$. Therefore,
using condition \eqref{med05}, we estimate
\begin{equation}
  \frac{\Delta \inro_{A}}{\Delta t}
  ~\sim~ \frac{1}{\Delta t} \cdot \frac{V^{2}}{\hbar^{2}}
    \: \tau_{B}^{2}
  ~=~ \frac{V^{2} \tau_{B}}{\hbar^{2}}
      \cdot \frac{\tau_{B}}{\Delta t}
  ~=~ \frac{1}{T_{A}} \cdot \frac{\tau_{B}}{\Delta t},
\label{med10} \end{equation}
as the integrals are nonzero within the interval of the length of
the order of $\tau_{B}$. If $\tau_{B} \ll \Delta t$ then the
correction is small (main contribution to the evolution of
$\inro_{A}$ is of the order of $1/T_{A}$, which is quite larger).
The key assumption that $\tau_{B} \ll \Delta t$ allows us
to conclude that the correlations between system ${\cal A}$
and ${\cal B}$ which were built before moment $t$ do not
significantly change the evolution of $\inro_{A}(t)$,
their influence is restricted to the moments from
a very short interval $(t, t-\tau_{B})$. New correlations,
within a much longer interval $(t,t+\Delta t)$, are building up
and have an effect on the evolution of $\inro_{A}(t)$. Initial
correlations have small significance and hence it is justified to
neglect them.

\subsubsection*{Discussion of the secular approximation}

Secular approximation consists in replacing the function
$F(\Omega\,'-\Omega)$ (defined in \eqref{me85}) in \eqref{me88}
by Kronecker delta, which leads to Eq.\eqref{me90}.
Our discussion of this replacement does not rise any doubts
when $|\Omega\,' - \Omega| \ll (\Delta t)^{-1}$,
because then $F(\Omega\,' - \Omega)$ is practically unity.
On the other hand for $|\Omega\,' - \Omega| \gg (\Delta t)^{-1}$
the function $F(\Omega\,' - \Omega)$ is practically zero.
The only problem is to justify the neglecting of the terms
for which $|\Omega\,' - \Omega| \sim (\Delta t)^{-1}$.

To explain this point, first use the fact that the free
evolution of matrix elements $\rho_{ab}$ is governed by
\begin{equation}
   \ddt \: \rho_{ab}^{(F)}
   ~=~ - \: \frac{i}{\hbar}
   \bra{a} \bigl[ H_{A}, \; \rho^{(F)} \:\bigr] \ket{b}
   ~=~ -i \omega_{ab} \rho_{ab}^{(F)},
\label{medd01} \end{equation}
where the small Lamb shift (due to $H_{LS}$) is ignored.
The solution is simple
\begin{equation}
   \rho_{ab}^{(F)}(t)
   ~=~ e^{-i \omega_{ab}t} \rho_{ab}^{(F)}(0).
\label{medd02} \end{equation}

Now, we analyze the dissipative term which is given by
\eqref{me120}. We do not discuss the nuances connected
with possible degeneracies. Obviously, we can write
\begin{equation}
    \rho_{mb} = \sum_{k} \delta_{bk} \rho_{mk}
    \qquad \mathrm{and} \qquad
    \rho_{am} = \sum_{k} \delta_{ak} \rho_{km},
\label{medd04} \end{equation}
which we use in \eqref{me120}. We also interchange
indices $k \leftrightarrow n$ in the second term and
similarly, in the third one we first interchange
$m \leftrightarrow n$ and then $k \leftrightarrow m$.
The result is
\begin{align}
& \hspace*{-8mm}
   \ddt \: \rho_{ab}(t) \Bigr|_{d.}
   ~=~ \sum_{m,n} \: \delta(\omega_{ma} - \omega_{nb}) \;
       K(am,bn) \; \rho_{mn}(t)
\nonumber \\
& \hspace*{30mm}
   -~\pol 
     \sum_{k,m,n} \: \delta_{bn} \:
     \delta(\omega_{mk} - \omega_{ak}) \;
     K(km,ka) \; \rho_{mn}(t)
\nonumber \\
& \hspace*{40mm}
   -~\pol 
     \sum_{k,m,n} \: \delta_{am} \:
     \delta(\omega_{bk} - \omega_{nk}) \;
     K(kb,kn) \; \rho_{mn}(t).
\label{medd07} \end{align}
We note that $\delta_{bn}$ implies $b=n$, and then $\omega_{nb}=0$.
Therefore
\begin{equation}
     \delta_{bn} \: \delta(\omega_{mk} - \omega_{ak})
   = \delta_{bn} \: \delta(\omega_{ma})
   = \delta_{bn} \: \delta(\omega_{ma} - \omega_{nb}),
\label{medd08} \end{equation}
since $\omega_{nb}=0$ and changes nothing. Similarly we have
\begin{equation}
     \delta_{am} \: \delta(\omega_{bk} - \omega_{nk})
   = \delta_{am} \: \delta(-\omega_{nb})
   = \delta_{bn} \: \delta(\omega_{ma} - \omega_{nb}).
\label{medd09} \end{equation}
Finally, we note that $ \delta(\omega_{ma} - \omega_{nb}) =%
\delta(\omega_{ab} - \omega_{mn})$, because Kronecker delta
is even. After these manipulations Eq. \eqref{medd07}
can be written as
\begin{align}
& \hspace*{-8mm}
   \ddt \: \rho_{ab}(t) \Bigr|_{d.}
   ~=~ \sum_{m,n} \: \delta(\omega_{ab} - \omega_{mn})
       \Bigl\{ K(am,bn)
   -\pol \: \delta_{bn} \: \sum_{k} K(km,ka)
\nonumber \\
& \hspace*{60mm}
   -\pol \: \delta_{am} \: \sum_{k} K(kb,kn)
   \Bigr\} \: \rho_{mn}(t).
\label{medd10} \end{align}
The expression in braces is denoted as $M_{am,bn}$
and we have
\begin{equation}
   \ddt \: \rho_{ab}(t) \Bigr|_{d.}
   ~=~ \sum_{m,n} \: \delta(\omega_{ab} - \omega_{mn})
       M_{am,bn} \: \rho_{mn}(t).
\label{medd13} \end{equation}
This is a specific form of ME, but useful in the discussion.
However, in the degenerate case some care must be exercised
and renewed considerations might be necessary.

We proceed to the discussion of conditions and/or possibilities
of neglecting the terms for which $|\Omega\,' %
- \Omega| \sim (\Delta t)^{-1}$. Interaction with the
reservoir certainly modifies the free evolution of
$\rho_{ab}^{(F)}(t)=e^{-i \omega_{ab}t} \rho_{ab}^{(F)}(0)$.
If Bohr frequencies of the oscillating elements
$\rho_{ab}$ i $\rho_{mn}$ are such that
$|\omega_{ab} - \omega_{mn}| \gg 1/T_{A}$,
then mutual couplings between these elements are quickly
averaged to zero (interfere destructively) before time $T_{A}$
elapses and before the interaction has enough time
to affect the evolution. In other words, if $|\omega_{ab} - %
\omega_{mn}|$ differs much from $1/T_{A}$ then the coupling
between corresponding matrix elements will have small
(weak) effect. This is the situation similar to the one
encountered in perturbation theory.
Namely, when the energies $|E_{a} - E_{b}| \gg V_{ab} = %
\elm{a}{V}{b}$ then the perturbation  has small (usually
negligible) effect.

Since, by assumption $T_{A} \gg \Delta t$ the discussed
situation corresponds, in fact, to the relation
$|\omega_{ab} - \omega_{cd}| \sim (\Delta t)^{-1}$.
This, in turn means, that such terms have little influence
on the evolution of the operator $\rho_{A}(t)$.
Such terms are neglected while passing from Eq. \eqref{me88}
to \eqref{me90}. Thus the last of our approximations is
justified.

\subsection{$V_{AB}=V^{\dagger}_{AB}$ -- nonhermiticity%
~of operators $A_{\alpha}$ i $X_{\alpha}$}

In our considerations we have adopted the interaction
hamiltonian between the system ${\cal A}$ and reservoir
${\cal B}$ in the form $V_{AB}=\sum_{\alpha} A_{\alpha} %
\otimes X_{\alpha}$, where operators $A_{\alpha}$ and
$X_{\alpha}$ do not have to be hermitian. Certainly, the full
interaction hamiltonian must be hermitian, so we conclude that
it must contain operators $A_{\alpha}$, $X_{\alpha}$
and their hermitian conjugates
$A_{\alpha}^{\dagger}$, $X_{\alpha}^{\dagger}$. Constructing
linear combinations we can always transform the interaction
hamiltonian into $V_{AB}=\sum_{\alpha} A_{\alpha}^{\,'} \otimes%
X_{\alpha}^{\,'}$, where the primed operators are hermitian.

We shall illustrate this with a simple example. Let the
interaction hamiltonian be of the form
\begin{equation}
   V_{AB}
   ~=~ A \otimes X^{\dagger} ~+~ A^{\dagger} \otimes X,
\label{mee03} \end{equation}
where operators $A$ and $X$ are nonhermitian, while the full
hamiltonian is clearly hermitian. We define new operators
\begin{align}
& \hspace*{-6mm}
   q = \frac{1}{\sqrt{2}} \bigl( A + A^{\dagger} \bigr),
   \hspace*{20mm}
   Q = \frac{1}{\sqrt{2}} \bigl( X + X^{\dagger} \bigr),
\nonumber \\
& \hspace*{-6mm}
   p = \frac{i}{\sqrt{2}} \bigl( A - A^{\dagger} \bigr),
   \hspace*{20mm}
   P = \frac{i}{\sqrt{2}} \bigl( X - X^{\dagger} \bigr),
\label{mee05} \end{align}
which are evidently hermitian. Expressing operators $A$, $X$
and their conjugates via the new ones, we obtain
\begin{align}
   V_{AB}
   &= \pol\bigl(q-ip\bigr)\otimes\bigl(Q+iP\bigr)
   ~+~ \pol\bigl(q+ip\bigr)\otimes\bigl(Q-iP\bigr)
\nonumber \\
   &= q \otimes Q ~+~ p \otimes P.
\label{mee06} \end{align}
This interaction hamiltonian is expressed as a sum of products
of hermitian operators. Hence construction of the interaction
hamiltonian with nonhermitian operators is allowed. One can
always build necessary combinations. However, in some
practical applications it is much more convenient to use
nonhermitian operators than the linear combinations.

\subsection{Vanishing average $\qav{X_{\alpha}}_{B}$}

In the main part of the lecture we assumed that Eq.\eqref{me37}
holds, that is the average $\qav{X_{\alpha}}_{B} \equiv %
\TTr_{B} \left\{ \: X_{\alpha} \, \rho_{B}(t) \: \right\} = 0$.
We have stated that it is not really restrictive.
We will show that it is true. This is so, because we can always
shift the energy scale. To see this, let us write
\begin{align}
  V^{\,'}_{AB}
  &= \sum_{\alpha}~A_{\alpha} \otimes
      \bigl( X_{\alpha} ~-~ \qav{X_{\alpha}}_{B} \bigr)
\nonumber \\
  &= \sum_{\alpha}~A_{\alpha} \otimes X_{\alpha}
  ~-~ \sum_{\alpha}~\qav{X_{\alpha}}_{B}
      \bigl( A_{\alpha} \otimes \idd_{B} \bigr),
\label{mef01} \end{align}
where $\qav{X_{\alpha}}_{B} = \TTr_{B}\{ \sib X_{\alpha}\}$
is a number not necessarily equal to zero. Then we have
\begin{equation}
  \qav{V^{\,'}_{AB}}_{B}
  ~=~ \sum_{\alpha}~A_{\alpha}
      \bigl( \qav{X_{\alpha}}_{B} ~-~ \qav{X_{\alpha}}_{B}
      \bigr) = 0,
\label{mef2} \end{equation}
which holds no matter whether numbers $\qav{X_{\alpha}}_{B}$
are zeroes or not. Full hamiltonian can then be written as
\begin{align}
   H_{AB}
   &= H_{A}\otimes\idd_{B}~+~\idd_{A}\otimes H_{B}~+~V_{AB}
\nonumber \\
   &= H_{A}\otimes\idd_{B}~+~\idd_{A}\otimes H_{B}
       ~+~V^{\,'}_{AB}
       ~+~ \sum_{\alpha}~\qav{X_{\alpha}}_{B}
           \bigl( A_{\alpha} \otimes \idd_{B}  \bigr)
\nonumber \\
   &= \bigl[ H_{A} ~+~ \sum_{\alpha}~\qav{X_{\alpha}}_{B}
         A_{\alpha} \bigr] \otimes \idd_{B}
       ~+~\idd_{A}\otimes H_{B}
       ~+~V^{\,'}_{AB}.
\label{mef3} \end{align}
Rescaled interaction term (the last one) has zero average
(as in \eqref{mef2}). This is achieved by the redefinition
of the energy scale in system ${\cal A}$ -- via redefinition
of the hamiltonian $H_{A}$. We conclude that the assumption
that the averages \eqref{me37} vanish is not really restrictive,
but simplifies the computation.

\subsection{Commutators of  operators $A_{\alpha}(\Omega)$}

In the main sections we have introduced the operators
$A_{\alpha}(\Omega)$ defined by relation \eqref{me45}.
The hamiltonian of system $\cal{A}$ is of the form
$H_{A} = \sum_{n} \hbar\omega_{n} \ket{n}\bra{n}$.
It is not difficult to find the commutator
$\bigl[ H_{A}, ~A_{\alpha}(\Omega) \bigr]$. Directly from
the definitions we obtain
\begin{align}
& \bigl[ H_{A}, ~A_{\alpha}(\Omega) \bigr]
  = \Bigr[\sum_{n} \hbar\omega_{n} \ket{n}\bra{n},
         ~\sum_{a,b} \delta(\omega_{ba} - \Omega) \:
          \ket{a}\elm{a}{A_{\alpha}}{b}\bra{b}
    \Bigl]
\nonumber \\
& \hspace*{10mm}
   = \sum_{a,b}
     \hbar (\omega_{a} - \omega_{b}) \:
     \delta(\omega_{ba} - \Omega)
     \ket{a} \elm{a}{A_{\alpha}}{b}\bra{b}
\nonumber \\
& \hspace*{10mm}
   = - \; \hbar \Omega
       \sum_{a,b}
       \delta(\omega_{ba} - \Omega)
       \ket{a} \elm{a}{A_{\alpha}}{b}\bra{b}
  ~=~ - \hbar \Omega A_{\alpha}(\Omega),
\label{meg03} \end{align}
which ends the calculation. Conjugation changes sign, so that
\begin{equation}
    \bigl[ H_{A}, ~A_{\alpha}^{\dagger}(\Omega) \bigr]
    ~=~ \hbar \Omega A_{\alpha}^{\dagger}(\Omega).
\label{meg04} \end{equation}
Heisenberg equation of motion follows from formula
\eqref{meg03}, and it is
\begin{equation}
   i \hbar \ddt A_{\alpha}^{(H)}(\Omega)
   ~=~ \bigl[ \, A_{\alpha}^{(H)}(\Omega), \; H_{A} \, \bigr]
   ~=~ \hbar \Omega A_{\alpha}^{(H)}(\Omega).
\label{meg05} \end{equation}
After integration we obtain $A_{\alpha}^{(H)}(\Omega)%
= e^{i\Omega t} A_{\alpha}(\Omega)$ which agrees with
\eqref{me55}. Finally, we present one more relation
\begin{align}
  & \bigl[ \,H_{A}, \;
            A^{\dagger}_{\alpha}(\Omega) A_{\beta}(\Omega)
    \: \bigr]
    = A^{\dagger}_{\alpha}(\Omega)
        \bigl[ \,H_{A}, \; A_{\beta}(\Omega) \: \bigr]
\nonumber \\
& \hspace*{50mm}
    ~+~ \bigl[ \,H_{A}, \; A_{\alpha}^{\dagger}(\Omega) \: \bigr]
        A_{\beta}(\Omega)
    =0,
\label{meg09} \end{align}
which follows immediately from the derived results.

\subsection{Explicit form of correlation functions%
~$\bar{G}_{\alpha \beta}(\tau)$}

Correlation function of the reservoir was defined in
\eqref{me62} or \eqref{me65f}. By assumption, reservoir
hamiltonian $H_{B}$ and the corresponding density operator
$\sib$ commute, so they have a common set of complete and
orthonormal eigenstates $\ket{z}$. Let us calculate the trace
in \eqref{me65f} in the chosen basis
\begin{align}
  \bar{G}_{\alpha \beta}(\tau)
  &= \TTr_{B} \left\{ \: \widetilde{X}_{\alpha}^{\dagger}(\tau) \:
         X_{\beta} \: \sib \: \right\}
  ~=~ \TTr_{B} \left\{ e^{ iH_{B}\tau/\hbar} \;
      X_{\alpha}^{\dagger} \;
               e^{-iH_{B}\tau/\hbar} \; X_{\beta} \:
      \sib \right\}
\nonumber \\
  &= \sum_{z,\xi}
      \elm{z}{e^{ iH_{B}\tau/\hbar} \: X_{\alpha}^{\dagger}\;
                   e^{-iH_{B}\tau/\hbar}}{\xi}
       \elm{\xi}{X_{\beta} \: \sib}{z}.
\label{meh02} \end{align}
In Eq.\eqref{me23b} we denoted the eigenvalues of $\sib$
by $p(z)$, hence
\begin{equation}
  \bar{G}_{\alpha \beta}(\tau)
    = \sum_{z,\xi} p(z)
      ~e^{ i\omega_{z\xi}\tau}
      \elm{z}{X_{\alpha}^{\dagger}}{\xi} \elm{\xi}{X_{\beta}}{z},
\label{meh03} \end{equation}
with $\omega_{z}=E_{z}/\hbar$,
and $\omega_{z\xi}=\omega_{z} - \omega_{\xi}$.

Expression \eqref{meh03} shows that the correlation function
$\bar{G}_{\alpha \beta}(\tau)$ is a complicated superposition
of functions which oscillate with Bohr frequencies $\omega_{z\xi}$.
Reservoir is assumed to be large, the discussed frequencies
are densely space (quasi-continuous).
If time $\tau$ is sufficiently large the oscillations interfere
destructively (average out to zero). We can expect that
reservoir correlation function decay quickly when time
$\tau=t_{1}-t_{2}$ increases. Characteristic decay time
is denoted by $\tau_{B}$  and assumed to be, by far, the shortest
time characterizing the system $\cal{A} + \cal{B}$.
When $\tau > \tau_{B}$ the correlation may be neglected.




\section{Summary}

In this summary we describe practical steps needed in the
construction of the ME for specified physical systems.

The first step consists in precise definition of the system
$\cal{A}$ and of the reservoir $\cal{B}$. We need to specify
their free hamiltonians $H_{A}$ and $H_{B}$ and (at least
sometimes) their eigenenergies and eigenstates.
Then we define the interaction  hamiltonian in the form
\begin{equation}
   V_{AB}
   ~=~ \sum_{\alpha} ~A_{\alpha} \otimes X_{\alpha}
   ~=~ \sum_{\alpha} ~A^{\dagger}_{\alpha}
       \otimes X^{\dagger}_{\alpha},
\label{mep01} \end{equation}
where $A_{\alpha},~X_{\alpha}$ are (correspondingly) operators
of system $\cal{A}$ and reservoir. We stress that these operator
need not be (separately) hermitian. It suffices that
the full interaction hamiltonian is hermitian.
We also need to specify the density operator $\sib$ describing
the state of the reservoir. It is worth remembering that
operator $H_{B}$ and $\sib$ commute. This implies that the
reservoir is in the stationary state. In the second step
of ME construction we build (identify) the following operators
\begin{equation}
   A_{\alpha}(\Omega)
   = \sum_{a,b} \delta(\omega_{ba} - \Omega) \:
     \ket{a} \elm{a}{A_{\alpha}}{b} \bra{b}.
\label{mep02} \end{equation}
The following matrix elements are computed in the third step
\begin{equation}
  W_{\alpha\beta}(\Omega)
  =  \int_{0}^{\infty} d\tau
     ~e^{i \Omega \tau} ~\bar{G}_{\alpha \beta}(\tau)
  = \int_{0}^{\infty} d\tau
     ~e^{i \Omega \tau}
     \: \TTr_{B} \bigl\{ \widetilde{X}^{\dagger}_{\alpha}(\tau) \,
     X_{\beta} \sib \bigr\}.
\label{mep03}
\end{equation}
They are seen to be partial Fourier transform of the reservoir
correlation functions. Reservoir operators are taken in the
interaction picture
\begin{equation}
   \widetilde{X}_{\alpha}(t)
   ~=~ e^{ iH_{B}t/\hbar} \; X_{\alpha} \; e^{-iH_{B}t/\hbar}.
\label{mep04} \end{equation}
Coefficients $W_{\alpha\beta}(\Omega)$ are then employed
to construct two hermitian matrices
\begin{align}
    \Gamma_{\alpha \beta}(\Omega)
    &= W_{\alpha\beta}(\Omega) ~+~ W^{\ast}_{\beta\alpha}(\Omega),
\nonumber \\
    \Delta_{\alpha \beta}(\Omega)
    & = \frac{1}{2i} \bigl[ \: W_{\alpha\beta}(\Omega)
      ~-~ W^{\ast}_{\beta\alpha}(\Omega) \: \bigr].
\label{mep05} \end{align}
We note that $\Gamma_{\alpha \beta}(\Omega)$ is a
positive-definite matrix and can be computed directly as
Fourier transform
\begin{equation}
   \Gamma_{\alpha \beta}(\Omega)
   = \int_{-\infty}^{\infty} d\tau ~e^{i \Omega \tau}
     ~\TTr_{B} \bigl\{ \widetilde{X}^{\dagger}_{\alpha}(\tau)
        X_{\beta} \sib \bigr\}
   = \int_{-\infty}^{\infty} d\tau ~e^{i \Omega \tau}
     \bar{G}_{\alpha \beta}(\tau).
\label{mep06a} \end{equation}
Parameters $\Gamma_{\alpha \beta}(\Omega)$, in practical
applications, are more important than $\Delta_{\alpha \beta}%
(\Omega)$. Explanation will be given later. The separate
expression for elements $\Delta_{\alpha\beta}(\Omega)$ is
\begin{equation}
  \Delta_{\alpha \beta}(\Omega)
  = \frac{1}{2i}
    \Bigl[
       \int_{0}^{\infty} d\tau ~e^{ i \Omega \tau}
      ~\TTr_{B} \bigl\{ \widetilde{X}_{\alpha}^{\dagger}(\tau)
       X_{\beta} \sib \bigr\}
   ~-~ \int_{0}^{\infty} d\tau ~e^{ - i \Omega \tau}
      ~\TTr_{B} \bigl\{ X_{\alpha}^{\dagger}
       \widetilde{X}_{\beta}(\tau) \sib \bigr\}
    \Bigr].
\label{mep6b} \end{equation}
Hence, calculation of coefficients $W_{\alpha\beta}(\Omega)$ can
be usually omitted.

Final construction of the proper ME is the fourth and the last
step. The above given quantities allow us to write the ME as
\begin{align}
& \ddt \, \rho_{A}(t)
  ~= -\:\frac{i}{\hbar}
       \bigl[ H_{A} + H_{LS}, \: \rho_{A}(t) \bigr]
\nonumber \\
& \hspace*{8mm}
   +~\frac{1}{\hbar^{2}}
     \sum_{\Omega} \sum_{\alpha,\beta}
     \Gamma_{\alpha \beta}(\Omega) \:
     \Big\{ \: A_{\beta}(\Omega) \: \rho_{A}(t) \:
               A^{\dagger}_{\alpha}(\Omega)
\nonumber \\
& \hspace*{45mm}
 -~ \pol \: 
     \Bigl[ A^{\dagger}_{\alpha}(\Omega) \:
            A_{\beta}(\Omega), \: \rho_{A}(t) \Bigr]_{+}
     \Bigr\},
\label{mep07} \end{align}
where the so-called Lamb-shift hamiltonian $H_{LS}$
is given as
\begin{equation}
   H_{LS}
   = \frac{1}{\hbar} \sum_{\Omega} \sum_{\alpha,\beta}
     \Delta_{\alpha \beta}(\Omega)
     A^{\dagger}_{\alpha}(\Omega) A_{\beta}(\Omega).
\label{mep08} \end{equation}
Energy shifts of the system $\cal{A}$ which are due to the
presence of $H_{LS}$ in the hamiltonian part, are usually
quite small and frequently negligible. This explains why
the role of matrix $\Delta_{\alpha \beta}$ is usually
less important than that of matrix $\Gamma_{\alpha \beta}$.

\vspace*{5mm}
\begin{center}
* * * * * * * * * * * * * * * * * * * * * * * * * * * * * *
\end{center}

\newpage





\begin{thebibliography}{99}
\bibitem{coh0}
    C. Cohen-Tannoudji, B. Diu, F. Lalo{\"{e}},
    {\em Quantum Mechanics},
    \\ Wiley-Interscience, New York 1991.
\bibitem{al}
    R. Alicki and K. Lendi, {\em Quantum Dynamical Semigroups and
    Applications},
    \\ Lect. Notes Phys 717, (Springer, Berlin Heidelberg 2007).
\bibitem{bp}
    H-P. Breuer, F. Petruccione,
    {\em The theory of open quantum systems},
    \\ Oxford University Press, 2002.
\bibitem{horn}
    K. Hornberger, {\em Introduction to decoherence theory},
    arXiv:quant-ph/0612118v2.
\bibitem{pres}
    J. Preskill, {\em Lecture notes on quantum computation}, \\
    http://www.theory.caltech.edu/$\sim$preskill/ph229.
\bibitem{coh}
    C. Cohen-Tannoudji, J. Dupont-Roc, G. Grynberg,
    {\em Atom--photon interactions},
    \\ Wiley, New York 1992.
\end{thebibliography}
\end{document}